\documentclass[12pt,a4paper]{elsarticle}
\usepackage[nottoc]{tocbibind}
\usepackage{graphicx}
\usepackage{slashed}
\usepackage{enumerate}
\usepackage{amsmath}
\usepackage{mathtools,cancel}
\usepackage[bottom]{footmisc}
\usepackage[scr=rsfso]{mathalfa}
\usepackage[hidelinks]{hyperref}
{

}
\usepackage{amsfonts}
\usepackage{psfrag}
\usepackage{color}
\usepackage{mathtools}
\usepackage{pgf,tikz}
\usepackage{mathrsfs}
\usetikzlibrary{arrows}
\usepackage{fancyhdr}
\usepackage{amssymb}
\usepackage{slashed}
\usepackage{enumitem}

\usepackage[T1]{fontenc}
\usepackage[utf8]{inputenc}

\newtheorem{remark}{Remark}

\newtheorem{proposition}{Proposition}[section]

\newtheorem{theorem}{Theorem}[section]
\newtheorem{corollary}{Corollary}[section]
\newtheorem{lemma}{Lemma}[section]

\numberwithin{equation}{section}
\allowdisplaybreaks[4]

\newenvironment{proof}{\smallskip\noindent\emph{Proof.}\hspace{1pt}}%
{\hspace{-5pt}{\nobreak\quad\nobreak\hfill\nobreak$\square$\vspace{8pt}%
		\par}\smallskip\goodbreak}

\newcommand{\be}{\begin{equation}}
\newcommand{\ee}{\end{equation}}

\newcommand{\bm}{\begin{align*}}
\newcommand{\enm}{\end{align*}}

\newcommand{\bespeq}{\begin{equation}\begin{split}}
\newcommand{\espeq}{\end{split}\end{equation}}

\newcommand{\tr}{\mbox{tr}}

\newcommand\restri[2]{{
		\left.\kern-\nulldelimiterspace 
		#1 
		\right|_{#2} 
}}

\definecolor{ffqqqq}{rgb}{1.,0.,0.}
\definecolor{uuuuuu}{rgb}{0.26666666666666666,0.26666666666666666,0.26666666666666666}

\makeatletter
\def\ps@pprintTitle{%
  \let\@oddhead\@empty
  \let\@evenhead\@empty
  \let\@oddfoot\@empty
  \let\@evenfoot\@oddfoot
}
\def\@author#1{\g@addto@macro\elsauthors{\normalsize%
    \def\baselinestretch{1}%
    \upshape\authorsep#1\unskip\textsuperscript{%
      \ifx\@fnmark\@empty\else\unskip\sep\@fnmark\let\sep=,\fi
      \ifx\@corref\@empty\else\unskip\sep\@corref\let\sep=,\fi
      }%
    \def\authorsep{\unskip,\space}%
    \global\let\@fnmark\@empty
    \global\let\@corref\@empty  
    \global\let\sep\@empty}%
    \@eadauthor={#1}
}
\makeatother

\begin{document}
\begin{frontmatter}
	
\title{Big-bang limit of $2+1$ gravity and Thurston boundary of Teichm\"uller space}
\author{Puskar Mondal\footnote{puskar\_mondal@fas.harvard.edu}
\\ Department of Mathematics, Harvard University,\\
Center of Mathematical Sciences and Applications, Harvard University}


\begin{abstract}
\vspace{5pt}
\par \noindent We study the asymptotic behavior of the solution curves of the dynamics of spacetimes of the topological type $\Sigma_{p}\times \mathbb{R}$, $p>1$, where $\Sigma_{p}$ is a closed Riemann surface of genus $p$, in the regime of $2+1$ dimensional classical general relativity. The configuration space of the gauge fixed dynamics is identified with the Teichm\"uller space ($\mathcal{T}\Sigma_{p}\approx \mathbb{R}^{6p-6}$) of $\Sigma_{p}$. Utilizing the properties of the Dirichlet energy of certain harmonic maps, estimates derived from the associated elliptic equations in conjunction with a few standard results of the theory of the compact Riemann surfaces, we prove that every non-trivial solution curve runs off the edge of the Teichm\"uller space at the limit of the big bang singularity and approaches the space of projective measured laminations/foliations ($\mathcal{PML}$ $\mathcal{PMF}$), the Thurston boundary of the Teichm\"uller space.
\end{abstract}
\end{frontmatter}

\begin{abstract}
 
\end{abstract}

\medskip



\section{Introduction}
\noindent 2+1 gravity formulated for spacetimes of the type $\Sigma_{p}\times \mathbb{R}$, where $\Sigma_{p}$ is the closed (compact without boundary) Riemann surface of genus $p>1$, is of considerable interest in mathematical relativity despite the fact that it does not allow for gravitational wave degrees of freedom and as such is devoid of straightforward physical significance. However, it becomes extremely important while studying `$3+1$' gravity on spacetimes of a certain topological type. \cite{moncrief1986reduction} studied Einstein's equations for  $U(1)$ symmetric vacuum spacetimes with spatial topology being a circle bundle over $\mathbb{S}^{2}$. Later \cite{choquet2001future, choquet2001future1, choquet2005topologically} studied the small data global well-posedness result for the vacuum Einstein equations on the spacetimes of the type $\mathbb{R}\times \mathfrak{P}_{p}$, where $\mathfrak{P}_{p}$ is a $\mathbb{S}^{1}$ bundle over a compact Riemann surface $\Sigma_{p}$ with genus $p$ strictly greater than $1$. Furthermore, the spacetime metric is assumed to be invariant with respect to the natural action of $U(1)$ along the bundle's circle fiber. This allowed $3+1$ gravity to be reducible to $2+1$ gravity coupled to a wave map that has the hyperbolic plane as its target space. In addition to these classical analyses, considerable attention has been paid to the corresponding quantum theory \cite{carlip2003quantum, witten19882,nelson19912+, ashtekar19892+}, where $2+1$ gravity is essentially treated as a toy model for $3+1$ quantum gravity. 

Despite such physical motivations to study $2+1$ gravity as a tool for studying physically interesting $3+1$ gravity, $2+1$ gravity is itself a mathematically rich topic with several open issues even at the purely classical level. A considerable amount of work has been done on purely classical $2+1$ gravity. Moncrief \cite{moncrief1989reduction} reduced the Einstein equations in $2+1$ dimensions to a Hamiltonian system over Teichm\"uller space, where the phase space of the dynamics was identified with the co-tangent bundle of Teichm\"uller space ($\approx \mathbb{R}^{12p-12}$). Later \cite{anderson1997global} proved the global existence of the Einstein equations on spacetimes of the topological type $\Sigma_{p}\times \mathbb{R}, p>1$ by controlling the Dirichlet energy (a proper function on Teichm\"uller space) of an associated harmonic map. Moncrief's extensive analysis of $2+1$ gravity (using constant mean curvature spatial harmonic gauge) in \cite{moncrief2007relativistic} led to several classical results of Teichm\"uller theory, which was obtained by means of purely relativistic/Riemannian geometric analysis. This included, e.g., the homeomorphism between the Teichm\"uller space and the space of holomorphic quadratic differentials (transverse-traceless tensors in the context of relativity), etc. In the same article, the term `\textit{Relativistic Teichm\"uller theory}' was coined. Through studying a Hamilton Jacobi equation whose complete solution determines all the solution curves of the reduced Einstein equations and a Monge-Ampere type equation which allows for a more explicit characterization of these solution curves, he defined a family of ray structures on the Teichm\"uller space of $\Sigma_{p}$. Studying the behavior of the associated Dirichlet energy, Moncrief \cite{moncrief2007relativistic} conjectured that \textit{each of these non-trivial solution curves runs off the edge of Teichm\"uller space at the limit of the big-bang singularity and attaches to the Thurston boundary of the Teichm\"uller space, that is, the space of projective measured laminations or foliations ($\mathcal{PML}$, $\mathcal{PMF}$)}. This, in principle, if holds true, then classifies the big bang singularities of `$2+1$' gravity as the points on the Thurston boundary and serves as another means to compactify Teichm\"uller space.

\cite{benedetti2001cosmological} studied the spacetimes of simplicial type (a dense subset in the space of all such spacetimes) in cosmological time gauge and obtained a similar result that the past singularity corresponds to the isometric action of the fundamental group of $\Sigma_{p}$ on a certain real tree. In other words, a point on the Thurston boundary is associated with the initial singularity. Later, based on the work of \cite{benedetti2001cosmological}, \cite{andersson2003constant} used barrier arguments to control the constant mean curvature slices relative to the cosmic time ones near the big bang singularities and thereby to show that Thurston boundary points are attained in the limit, by the former as well as the latter. Despite the fact that these results conform to the conjecture of Moncrief to a large extent, they lack direct arguments and also differ in the choice of gauge. Whether this result is gauge invariant is currently unknown. Therefore, it is worth proving the conjecture by direct analysis of the Einstein evolution and constraint equations in the CMCSH gauge.

In addition to the general relativistic perspective, M. Wolf \cite{wolf1989teichmuller} established the homeomorphism between the space of holomorphic quadratic differential and the Teichm\"uller space of $\Sigma_{p}$ by utilizing the complex analytic machinery such as the Beltrami differential (stretching) of the associated harmonic map. One may naively expect that Wolf's result might be directly applicable to the relativistic case since the transverse-traceless tensor of GR may be associated with a holomorphic quadratic differential. However, in Wolf's case, the domain is kept fixed while the dynamics occur on the target surface and therefore the available machinery from complex analysis became useful. But, in the relativistic case, the domain (conformal structure) varies while the target is fixed (an interior point of the Teichm\"uller space). Therefore, the traditional machinery becomes useless and we are left with tools that are only seemingly accessible through GR.  

In this article, we aim to study the '$2+1$' gravity on vacuum spacetimes of topological type $\Sigma_{p}\times \mathbb{R}$ in constant mean extrinsic curvature spatial harmonic gauge (CMCSH). Utilizing the direct estimates from the Einstein evolution and constraint equations in conjunction with a few established results from \cite{moncrief2007relativistic} and the theory of Riemann surfaces, we show via a direct argument that indeed Moncrief's conjecture holds true, that is, at the limit of the big-bang singularity, the conformal geometry degenerates and every corresponding non-trivial solution curve attaches to the Thurston boundary. 

 {Essentially, the physical universe volume collapses in the limit of big bang singularity by its very definition. Therefore, interesting dynamics, if any, are expected to be observed in an appropriate conformal setting. My main theorems deal with the asymptotic behavior of the conformal dynamics. Let us denote a Lorentzian spacetime by $(\widetilde{M},\widetilde{g})$ which in my context is diffeomorphic to $\Sigma_{p}\times \mathbb{R}$, where $\Sigma_{p}$ is a compact Riemann surface with genus $p>1$ and we are interested in Lorentzian metrics $\widetilde{g}$ up to diffeomorphisms of $\widetilde{M}$ verifying vacuum Einstein's equations 
\begin{eqnarray}
\label{eq:eom}
\textcolor{blue}{\text{Ric}}[\widetilde{g}]=0.   
\end{eqnarray}
The initial value problem of Einstein's equations consists of prescribing a pair $(g^{0}_{ij},K^{0}_{ij})$ on a space-like hypersurface (diffeomorphic to $\Sigma_{p}$) embedded in the spacetime that verifies the so-called constraint equations (see equations \ref{eq:HC} and \ref{eq:momentum}). Here $g^{0}_{ij}$ and $K^{0}_{ij}$ are an induced Riemannian metric on the Cauchy hypersurface and the associated second fundamental form, respectively. A solution of Einstein's evolution equation in an appropriate gauge given initial data $(g^{0}_{ij},K^{0}_{ij})$ is map $(g^{0}_{ij},K^{0}_{ij}) \mapsto (g_{ij}(t),K_{ij}(t),N(t),X^{i}(t))$ that verifies the axioms of the Cauchy problem i.e., existence, uniqueness (up to diffeomorphism), and continuity (or Cauchy stability). Here $N$ and $X^{i}$ are the lapse function and the shift vector field that together with the metric $g_{ij}$ completely determines the spacetime metric (see equation \ref{eq:spacetime}). In dimensions $n\geq 2$ there is a well-known technique pioneered
by Lichnerowicz, for solving the constraint equations on a constant-mean curvature hypersurface (see \cite{choquet2009general}, Bartnik and Isenberg \cite{constraint1} for detailed expositions of this `conformal' method). In the context of $2+1$ gravity, every Riemannian metric $g$ on $\Sigma_{p}$ is uniquely globally (pointwise) conformal to a metric $\gamma$ that verifies $R(\gamma)=-1$. In this case every Riemannian metric $g$ on $\Sigma_{p}$ can be uniquely expressed as $g_{ij}=e^{2\varphi}\gamma_{ij}$ for $\varphi:\Sigma_{p}\to \mathbb{R}$. More accurately, one may equivalently consider the data $(\gamma,k^{TT},\tr_{g}k,e^{\varphi})$ where $K_{ij}=k^{TT}_{ij}+\frac{\tr_{g}k}{2}g_{ij}$, $k^{TT}$ is transverse-traceless with respect to $g$ i.e., $g^{ij}k^{TT}_{ij}=0, \nabla[g]^{j}k^{TT}_{ij}=0$, $k^{TT}_{ij}$ does not scale with $\varphi$ while $(k^{TT})^{ij}$ does in our convention (see section \ref{reduced}). We work in constant mean extrinsic curvature gauge or CMC gauge throughout (i.e., $\tau:=\tr_{g}K=\text{function of time alone}$) in which the constraints (Hamiltonian and momentum constraints \ref{eq:HC}-\ref{eq:momentum}) decouple and more precisely the scalar field $\varphi$ verifies the so-called Lichnerowicz equation (\ref{eq:constraint}). The first of the three main theorems regarding the asymptotics of conformal dynamics is stated below. Even though this first theorem can be deduced from \cite{moncrief2007relativistic} with little work, we provide it for completeness since it plays an important role in the other main two theorems that we prove in this article.}

\begin{theorem}
\label{theorem1}
 {Let $\Sigma_{p}$ be a closed (compact without boundary) Riemann surface of genus $p>1$ and $(\Sigma_{p}\times I,\widetilde{g}), I\subset \mathbb{R}$ be a globally hyperbolic spacetime solving vacuum Einstein equations (\ref{eq:eom}), which is the maximally globally hyperbolic development of the initial data $(g_{0},K_{0})$ on $\Sigma_{p}$ with mean curvature $\tau_{0}=tr_{g_{0}}K_{0}<0$ (or equivalently the data $(\gamma,k^{TT}_{0},\varphi_{0},\tau_{0}=\tr_{g_{0}}K_{0}), R(\gamma)=-1$, $\varphi_{0}$ verifies Lichnerowicz equation (\ref{eq:constraint}) on $\Sigma_{p}$).  Every such non-trivial solution asymptotically approaches the fixed point solution ($R(\gamma)=-1, k^{TT}=0, N=2, X^{i}=0$) of the dynamical equations in the limit $\tau:=\tr_{g}K\to 0$ (future) and every such non-trivial solution curve runs off the edge of the configuration space (Teichm\"uller space) in the limit of the big-bang singularity $\tau\to-\infty$ (past).}   
\end{theorem}

 {The proof of this theorem crucially depends on the `in time' behavior of the Dirichlet energy of an appropriately constructed harmonic map (discussed in detail in section \ref{harmonic}). Such Dirichlet energy turns out to be a proper function on the configuration space or the Teichm\"uller space i.e., the inverse images of compact sets are compact. Therefore, the proof of the future asymptotic behavior amounts to proving the uniform boundedness of the Dirichlet energy along any solution curve (more precisely, the range of the Dirichlet energy should be a compact set of the positive real line). Proving this property in turn requires several estimates on the scalar field $\varphi$, the lapse function $N$, and the transverse-traceless momentum $k^{TT}$. In addition, we  also need to establish the existence of a solution to the conformal system and construct a solution curve on the configuration space (Teichm\"uller space). We accomplish this by means of the ray structure defined by Moncrief \cite{moncrief2007relativistic} and a Hamilton-Jacobi analysis. With the help of the estimates proven in lemma \ref{scalarestimate}, lemma \ref{momentumestimate}, and corollary \ref{lapseestimate}, we prove the uniform boundedness of the Dirichlet energy along any future-directed solution curve and that its range is a compact set on the positive real line. The second part of the theorem follows in a similar way but concerns solving Einstein's equations in the reversed time. In this case, the main task is to prove the blow-up of the Dirichlet energy as one approaches the big bang i.e., $\tau:=\tr_{g}K\to-\infty$. With the estimates proven in lemma \ref{scalarestimate}, lemma \ref{momentumestimate}, and corollary \ref{lapseestimate}, we provide a bound on the blow-up rate of the Dirichlet energy as $\tau\to-\infty$.} 

 {In order to characterize the space of big-bang singularities, it is essential to study the behavior of certain intrinsic properties associated with $\Sigma_{p}$ at the limit $\tau:=\tr_{g}K\to-\infty$. In particular, we  want to study the holomorphic quadratic differential  (defined in section \ref{HQD}) corresponding to the transverse-traceless tensor $k^{TT}$ and the associated measured laminations/foliations (defined in \ref{teichmuller}). In the next theorem, we prove the asymptotic behavior of the relationship between the hyperbolic length of a closed geodesic in a homotopy class and its transverse measure with respect to the horizontal and vertical foliations defined by the transverse-traceless tensor $k^{TT}$ (or the equivalent holomorphic quadratic differential). More precisely, the ratio of hyperbolic length and transverse measure against the vertical foliation approaches a constant independent of the homotopy class of curves as the big bang is approached. The corresponding horizontal measure approaches zero at the same limit. Notice that this behavior of the horizontal and vertical measured foliations is observed at the large limit of the Dirichlet energy of suitably defined harmonic maps in \cite{minsky1992harmonic}. However one ought to be cautious in a direct analogy. Below is the second theorem that we shall prove in this article. $\pi_{1}(\Sigma_{p})$ is used to denote the fundamental group of $\Sigma_{p}$ based at any point $x\in\Sigma_{p}$ \footnote{Choice of basepoint does not matter since changing basepoint results in changing conjugacy class}.}

\begin{theorem}
 {Let $\Sigma_{p}$ be a closed (compact without boundary) Riemann surface of genus $p>1$ and the conformal data $(\gamma,k^{TT},\tau,e^{\varphi},N,X)$ defined by the solution of the Gauss map equation (\ref{eq:gauss}), Lichnerowicz equation (\ref{eq:constraint}), and the elliptic equations (\ref{eq:lapse}-\ref{eq:shift}) solve the gauge fixed Einstein equations via the associated Hamilton-Jacobi equation (\ref{eq:HJ}) \footnote{Essentially a solution to the gauge fixed vacuum Einstein's equations}. The ratio of the transverse measure of any non-trivial element $\mathcal{C}$ of $\pi_{1}(\Sigma_{p})$ with respect to the vertical measured foliation of the natural holomorphic quadratic differential $\phi:=(k^{TT}_{11}-\sqrt{-1} k^{TT}_{12})dz^{2},$\footnote{$z$ is the usual complexification of the real coordinates i.e., in local chart $z=x+\sqrt{-1}y$} and its hyperbolic length that is the length with respect to the metric $\gamma$ approaches to a finite constant independent of any homotopy class in the limit of the big-bang singularity i.e., $\tau\to-\infty$ along every sequence on the solution curve. The transverse measure with respect to the horizontal foliation associated with the holomorphic quadratic differential $\phi$ collapses to zero in the same limit.}    
\end{theorem}

 {In order to prove this theorem, first we prove three lemmas \ref{theorem2proof1}, \ref{theoremproof2}, and \ref{theorem2proof3}. In the first lemma \ref{theorem2proof1}, we prove the boundedness of the transverse-traceless tensor. However, as it turns out, this is not enough to establish the theorem. In fact, it is possible to directly prove that the diffeomorphism invariant entity $|k^{TT}|^{2}_{\gamma}:=\gamma^{ik}\gamma^{jl}k^{TT}_{ij}k^{TT}_{kl}$ is uniform over $\Sigma_{p}$ as one asymptotically approaches big-bang. This is accomplished in lemma \ref{theoremproof2} by solving the evolution equation for $|k^{TT}|^{2}_{\gamma}$, the Lichnerowicz equation (\ref{eq:constraint}), the equations for the lapse function (\ref{eq:lapse}) and the shift vector field (\ref{eq:shift}) in the limit $\tau\to-\infty$. Such a uniform estimate then allows me to relate the hyperbolic length of a closed geodesic in a particular homotopy class to its measure against the specific measured foliation (described by the holomorphic quadratic differential through Hubbard-Masur homomorphism, see section \ref{HQD}) and this relation does not depend on the homotopy class of curves. This is accomplished by analyzing Moncrief's \cite{moncrief2007relativistic} Gauss map equation (\ref{eq:gauss}) in lemma \ref{theorem2proof3}.  The proof of theorem 2 then follows in a straightforward way. The final piece that remains is to characterize the asymptotic state. More precisely, since in theorem 1, we have established that every solution curve runs off the edge of the configuration space at the big-bang limit, the question remains whether such rays limit so a nice space that serves as the boundary of the configuration space. Using the result of theorem 2, we prove that every solution curve indeed approaches the Thurston boundary (also known as the space of projective laminations ($\mathcal{PML}$) or foliations ($\mathcal{PMF}$) of the Teichm\"uller space in the limit. Therefore the space of big bang singularities is a subset of the Thurston boundary of the Teichm\"uller space. One unanswered question that remains is proving that the space of big-bang singularity is exactly the Thurston boundary i.e., each point on the Thurston boundary can be realized as a limit of a unique solution curve as $\tau\to-\infty$. We intend to investigate this in the future since currently, we are unable to do so with the available methods. Below is the final theorem that establishes the convergence of a solution sequence leaving every compact set of the Teichm\"uller space to the Thurston boundary at the limit $\tau\to-\infty$.}          

\begin{theorem}
 {Every non-trivial solution curve of the reduced Einsteinian dynamics runs off the edge of the Teichm\"uller space at the limit of big-bang singularity and attaches to the Thurston boundary of the Teichm\"uller space, that is, the space of projective measured laminations or foliations ($\mathcal{PML}$, $\mathcal{PMF}$)}   
\end{theorem}

\begin{center}
\begin{figure}
\label{penrose}
\begin{center}
\includegraphics[scale=0.15]{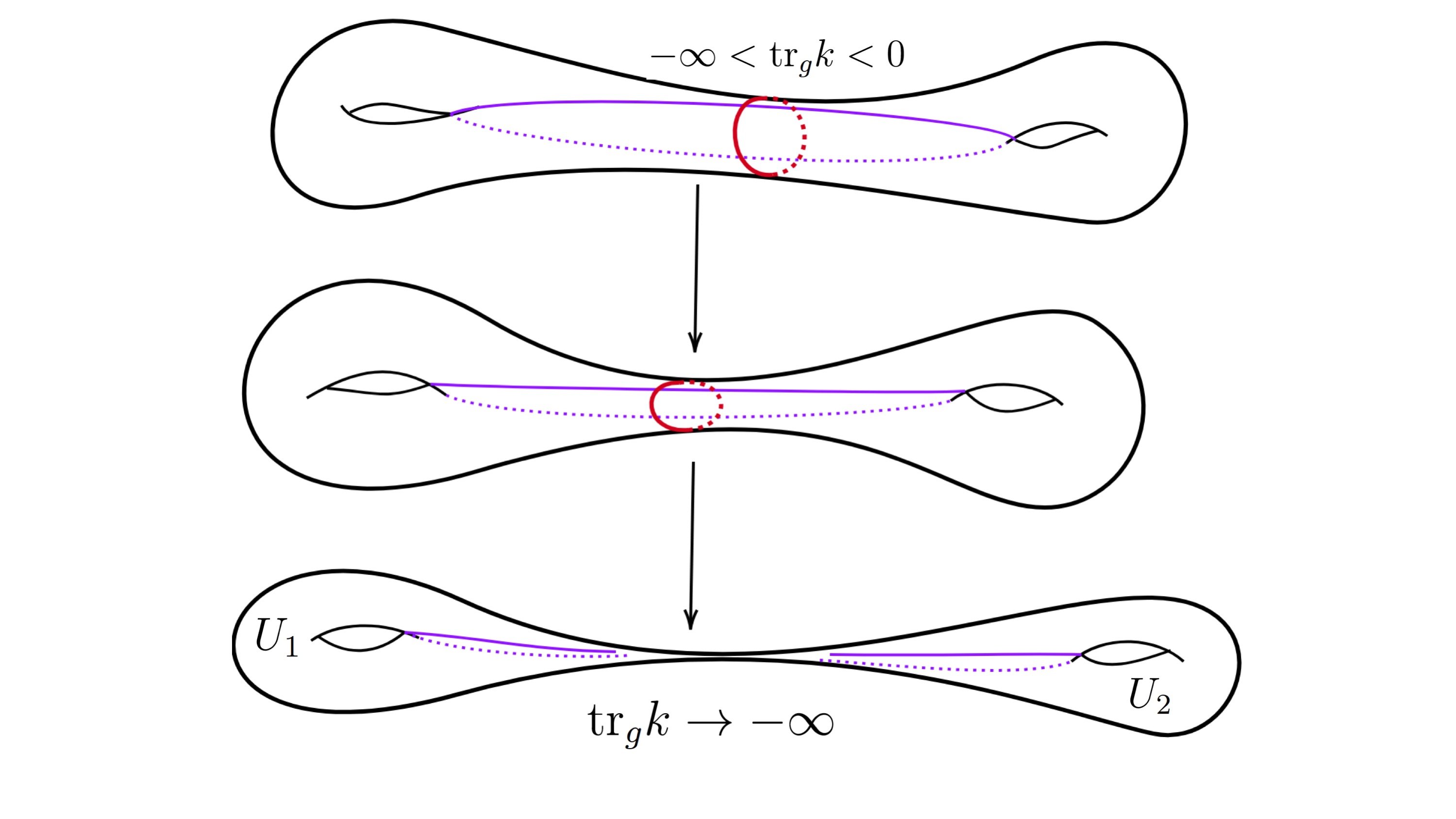}
\end{center}
\begin{center}
\caption{The schematics of the conformal dynamics on the configuration space $\mathcal{T}\Sigma_{p}$ ($\approx \mathbb{R}^{6p-6}$). As the big bang ($\tau\to-\infty$) approaches, the physical universe ($\Sigma_{2}$ in the figure) degenerates along a homotopically non-trivial geodesic (red) to produce two connected components $U_{1}$ and $U_{2}$ in the limit. This could also be interpreted as the emergence of the physical universe through the coalescence of $U_{1}$ and $U_{2}$ if one moves away from the big bang in time.}
\label{fig:pdf}
\end{center}
\end{figure}
\end{center}

An interesting physical consequence of the result is a different interpretation of the big-bang singularity. Clearly, the big bang is a crushing singularity in the sense that the volume of the physical universe approaches zero. However, if one extracts the conformal structure, then we see that my result indicates a degeneration of the physical universe along homotopically non-trivial loops in the big-bang limit (approaching the Thurston boundary of the configuration space). In other words, one obtains a set of connected components that make up the original Riemann surface of the physical universe (see figure \ref{penrose}). If one runs the time forward from this big-bang limit, then the degeneration phenomenon is reversed or one observes the self-gravitating of the connected components to constitute the physical universe. Physically, therefore, the $2+1$ dimensional big bang can be interpreted as the emergence of a single physical universe through the coalescence of multiple universes even though there is no `gravity' in the vacuum of $2+1$ dimensions. Another important question to ask would be the stability of the big-bang singularity and therefore of this degeneration process. 


The structure of the article is as follows. We begin with introducing the necessary background for the theory of Riemann surfaces such as harmonic maps, holomorphic quadratic differentials, the associated measured foliations, and their transverse measures etc. Then we study the reduced Einstein equations through a conformal technique and obtain the estimates necessary from the associated elliptic PDEs. Finally, we state the relativistic interpretation of the concepts from Riemann surface theory and prove using the estimates obtained that the conjecture holds true, that is, at the limit of big-bang singularity, every non-trivial solution curve runs off the edge of the Teichm\"uller space and attaches to the space of projective measured foliations/laminations and exhausts these spaces. We conclude by discussing the potential validity of the conjecture with the inclusion of a cosmological constant and suitable matter sources.   

\section{\textcolor{blue}{Preliminaries}}
\noindent We denote the `$2+1$' spacetime by $\widetilde{M}$ with its topology being $\Sigma_{p}\times \mathbb{R}$. Here, $\Sigma_{p}$ is a closed (compact without boundary) Riemann surface with genus $p>1$. The space of Riemannian metrics on $\Sigma_{p}$ is denoted by $\mathcal{M}$ and its closed submanifold $\mathcal{M}_{-1}$ is defined as follows 
\begin{eqnarray}
\mathcal{M}_{-1}=\{\gamma\in \mathcal{M}|R(\gamma)=-1\},
\end{eqnarray}
where $R(\gamma)$ is the scalar curvature of the metric $\gamma$. The space of symmetric 2-tensor fields is denoted by $S^{0}_{2}(\Sigma_{p})$. The $L^{2}$ inner product with respect to the metric $\gamma\in \mathcal{M}$ between any two elements $A$ and $B$ of $S^{0}_{2}(\Sigma_{p})$ is defined as 
\begin{eqnarray}
\langle A,B\rangle_{L^{2}}:=\int_{\Sigma_{p}}A_{ij}B_{kl}\gamma^{ik}\gamma^{jl}\mu_{\gamma},
\end{eqnarray}
where $\mu_{\gamma}=\sqrt{\det(\gamma_{ij})}dx^{1}\wedge dx^{2}$ is the volume form on $\Sigma_{p}$. Abusing notation we will use $\mu_{\gamma}$ for both $\sqrt{\det(\gamma_{ij})}$ and the volume form. Unless otherwise stated, we will consider an element of $\mathcal{M}$ in isothermal coordinates that is $\mathcal{M}\ni\gamma:=e^{\eta(z)}|dz|^{2},\eta: \Sigma_{p}\to \mathbb{R}$. The Laplacian $\Delta_{\gamma}$ is defined so as to have non-negative spectrum on $\Sigma_{p}$, that is, $\Delta_{\gamma}:=-\gamma^{ij}\nabla_{i}\nabla_{j}$. By a nontrivial element of $\pi_{1}(\Sigma_{p})$, we will always mean a non-trivial closed curve since there is a one-to-one correspondence between the homotopy classes of essential (not homotopic to a point or neighborhood of a puncture) closed curve and the conjugacy classes of non-trivial elements in $\pi_{1}(\Sigma_{p})$. The group of diffeomorphisms (of $\Sigma_{p}$) and its identity component are denoted by $\mathcal{D}$ and $\mathcal{D}_{0}$, respectively.

\subsection{Background on Teichm\"uller space}
\label{teichmuller}
\noindent Teichm\"uller space is studied from an algebraic topologic perspective \cite{farb2011primer, imayoshi2012introduction}, a complex analytic perspective \cite{nag1988complex, imayoshi2012introduction}, and a Riemannian geometric perspective\cite{tromba2012teichmuller}. Here, we will focus mainly on the latter as the Teichm\"uller space while viewed from a Riemannian geometric perspective naturally appears as the configuration space of vacuum Einstein gravity (with or without a positive cosmological constant) on $\Sigma_{p}\times R$. Nevertheless, we will state the algebraic topologic definition of Teichm\"uller space and show how this is connected to Einstein's gravity. 
 The Teichm\"uller space of $\Sigma_{p}$ is defined as the space of homomorphisms (more accurately the discrete and faithful representations) of the fundamental group of $\Sigma_{p}$ into the isometry group of its universal cover that is the hyperbolic plane modulo the action of the isometry group by conjugation. If the Poinca\'re disk model of the hyperbolic plane is assumed, then the \textcolor{blue}{Teichm\"uller space can be defined to be} 
 \begin{eqnarray}
 \mathcal{T}\Sigma_{p}&:=&\rho(\pi_{1}(\Sigma_{p}), PSL_{2}\mathbb{R})/PSL_{2}\mathbb{R} \text{conj}\\\nonumber
 &&\subset Hom(\pi_{1}(\Sigma_{p}), PSL_{2}\mathbb{R})/PSL_{2}\mathbb{R} \text{conj},\nonumber
 \end{eqnarray}
where $\rho$ denotes the space of discrete and faithful representations (sometimes representations abusing notation) and $PSL_{2}\mathbb{R}\text{conj}$ denotes conjugation operation by elements of $PSL_{2}\mathbb{R}$. Dimension of $\mathcal{T}\Sigma_{p}$ may be calculated as follows. The space of representations $\rho(\pi_{1}(\Sigma_{p}), PSL_{2}\mathbb{R})$ is moded out by the $PSL_{2}\mathbb{R}$ conjugation so as to remove the base point of the homotopy (at the level of loops). This definition precisely identifies the ways to equip $\Sigma_{p}$ with distinct conformal structures (or hyperbolic structures). The fundamental group $\pi_{1}(\Sigma_{p})$ is to be viewed as a discrete and faithful subgroup of $PSL_{2}\mathbb{R}$ and as such is finitely generated ($2p$ generators). The dimension of $PSL_{2}\mathbb{R}$ is 3 and action by conjugation by an element of $PSL_{2}\mathbb{R}$ produces equivalence classes (with respect to gauge transformation in physics terminology). In addition, the generators $(A_{i},B_{i})_{i=1}^{p}$ satisfy the commutation relation $\prod_{i=1}^{p}A_{i}B_{i}A^{-1}_{i}B^{-1}_{i}=id$ implying the representation $\rho \in Hom(\pi_{1}(\Sigma_{p}),PSL_{2}\mathbb{R})/PS_{2}R~conj$ would satisfy $\prod_{i=1}^{p}\rho(A_{i})\rho(B_{i})\rho(A_{i})^{-1}\rho(B_{i})^{-1}=id$ as well. Therefore we lose $3+3=6$ degrees of freedom out of $2g\times3=6p$ and the dimension of the Teichmuller space turns out to be $6p-6$. Let us now show how this is related to vacuum Einstein dynamics. The vacuum Einstein equations in $2+1$ dimension read 
\begin{eqnarray}
\label{eq:einf}
R_{\mu\nu}=0,
\end{eqnarray}
where $(\mu,\nu)$ correspond to the spacetime indices. Now, in $2+1$ dimension, the vanishing of the Ricci tensor ($R_{\mu\nu}$) implies the vanishing of the full Riemann tensor (or the sectional curvature) and therefore, the solutions of the Einstein equations are necessarily the flat spacetimes and consequently are locally isometric to the Minkowski spacetime. Now we are interested in flat spacetimes foliated by $\Sigma_{p}$. In order to obtain the solution space, I, therefore, need to identify the space of homomorphisms (space of discrete and faithful representations to be precise) of $\pi_{1}(\Sigma_{p}\times R)$ into the isometry group of the flat spacetimes, which in this case is the full Poincare group $ISO(2,1)$. Now $\pi_{1}(\Sigma_{p}\times R)\approx \pi_{1}(\Sigma_{p})$ and therefore the solution space is described as
\begin{eqnarray}
Ein_{S}&=& \rho(\pi_{1}(\Sigma_{p}), ISO(2,1))/ISO(2,1) \text{conj},
\end{eqnarray}
where $Ein_{S}$ is the space of solutions of the equation (\ref{eq:einf}). 
In a similar way, we may compute the dimension of $Ein_{S}$. Note that now the isometry group $ISO(2,1)$ has dimension $6$ and therefore following the exact same procedure, we obtain the dimension of $Ein_{S}$ to be $12p-12$. Therefore, the full solution space is twice the dimension of the Teichm\"uller space. One immediate guess would be that the co-tangent bundle $T^{*}\mathcal{T}\Sigma_{p}$ of the Teichm\"uller space acts as the full solution space, which is precisely the case as shown in \cite{moncrief1989reduction, moncrief2007relativistic}. $T^{*}\mathcal{T}\Sigma_{p}$ is indeed the phase space of the reduced dynamics. We will get back to this point in detail later (see section \ref{reduced}). Let us first develop the concepts of geodesic currents, measured laminations ($\mathcal{ML}$), and foliations ($\mathcal{MF}$), which will be required to prove the conjecture.   

Let us now introduce a few elementary concepts from the theory of Riemann surfaces. From elementary hyperbolic geometry, we know that there exists a unique geodesic between any two distinct points lying on the boundary of the Poinca\'re disc (in this model of the hyperbolic 2-space). Therefore, we define the set of all un-oriented geodesics on $\tilde{\Sigma}_{g}$ (lift of $\Sigma_{p}$ to its universal cover) as the $\mathbb{Z}_{2}$ graded double boundary of $\tilde{\Sigma}$ i.e., 
$G(\tilde{\Sigma}_{g})$=$\{$The set of all un-oriented geodesics on $\tilde{\Sigma}\}$ $\approx (\mathbb{S}^{1}_{\infty}\times \mathbb{S}^{1}_{\infty}-\Delta)/\mathbb{Z}_{2}$, where $\Delta$ represents the diagonal. A geodesic current is a radon measure on $G(\tilde{\Sigma})$ which is invariant under the $\pi_{1}(\Sigma_{p})$ action (see \cite{lustig2019north, bridgeman2007distribution} for more details and see \cite{simon1983lectures} for details about radon measures). The property of a radon measure that would be of particular interest to us is that it is locally finite. In a sense, a geodesic current is essentially an assignment of a radon measure to the open sets of $G(\tilde{\Sigma})$, which remain invariant under the action of the fundamental group $\pi_{1}(\Sigma_{p})$. This $\pi_{1}(\Sigma_{p})$ invariance property of the geodesic currents allows one to define it on the space of geodesics on $\Sigma_{p}$ i.e., $G(\Sigma_{p})=G(\tilde{\Sigma}_{g})/\pi_{1}(\Sigma_{p})$ (note that the action of $\pi_{1}(\Sigma_{p})$ extends continuously to $\partial \tilde{\Sigma}_{g}$). Now, for a closed hyperbolic surface of genus greater than 1, $\pi_{1}(\Sigma_{p})$ while viewed as a proper discrete subgroup of the isometry group of the hyperbolic plane that is $PSL_{2}\mathbb{R}$, consists of hyperbolic (also called loxodromic) elements only (see \cite{farb2011primer, ratcliffe1994hyperbolic} for a detailed classification of the types of isometries of $\mathbb{H}^{2}$). Each element of $\pi_{1}(\Sigma_{p})$ has an axis geodesic along which it acts by translation and in general it has two fixed points: one attracting, one repelling. Therefore each element of $\pi_{1}(\Sigma_{p})$, a homotopy class of nontrivial loops (rectifiable), has a unique geodesic representative. Whenever we will consider the length of a non-trivial closed curve on $\Sigma_{p}$ we will always mean the length of the geodesic in its homotopy class. A geodesic lamination is a closed subset of $\Sigma_{p}$ which is the union of disjoint geodesics. A measured lamination is defined as a geodesic lamination equipped with a transverse measure (invariant under translations along the leaves of the lamination). Clearly, the space of measured laminations is a subset of the space of geodesic currents. A geodesic foliation may be thought of as the union of the geodesics which are also integral curves of a vector field. Zeros of the vector field correspond to the singularities of the foliation. One may similarly assign a transverse measure to the foliation promoting it to a measured foliation. There is a natural homeomorphism between the space of measured laminations and measured foliations (via a straightening map; see Fig [\ref{fig:foli}]). This homeomorphism persists at the level of corresponding projective spaces. This projective space (projective measured laminations or foliations) is the Thurston boundary of the Teichm\"uller space. At this point, it suffices to know this fact and therefore, we do not dwell on this matter further but rather provide a small detail in the appendix. Interested readers are referred to the same.  
  
\begin{center}
\begin{figure}
\begin{center}
\includegraphics[scale=0.4]{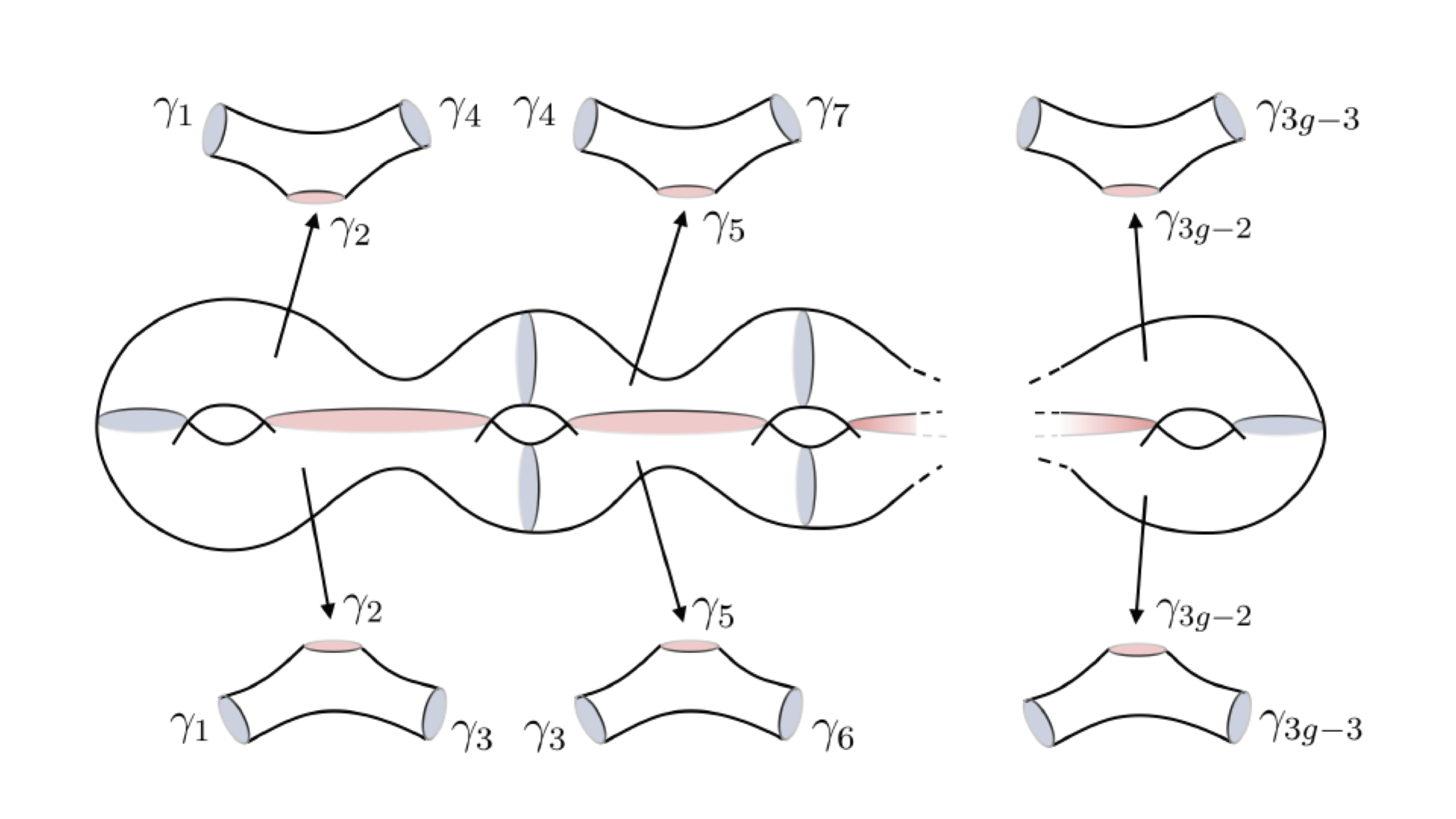}
\end{center}
\begin{center}
\caption{Pants decomposition of the hyperbolic surface $\Sigma_{p}$: hyperbolic length of $\gamma_{i}$ together with the twist about the same geodesic $\gamma_{i}$ parametrizes the Teichm\"uller space.}
\label{fig:fenchel}
\end{center}
\end{figure}
\end{center}

\subsection{Homeomorphism between $\mathcal{ML}$, $\mathcal{MF}$, and $\mathcal{QD}$}
\label{HQD}
\begin{center}
\begin{figure}
\begin{center}
\includegraphics[scale=0.1]{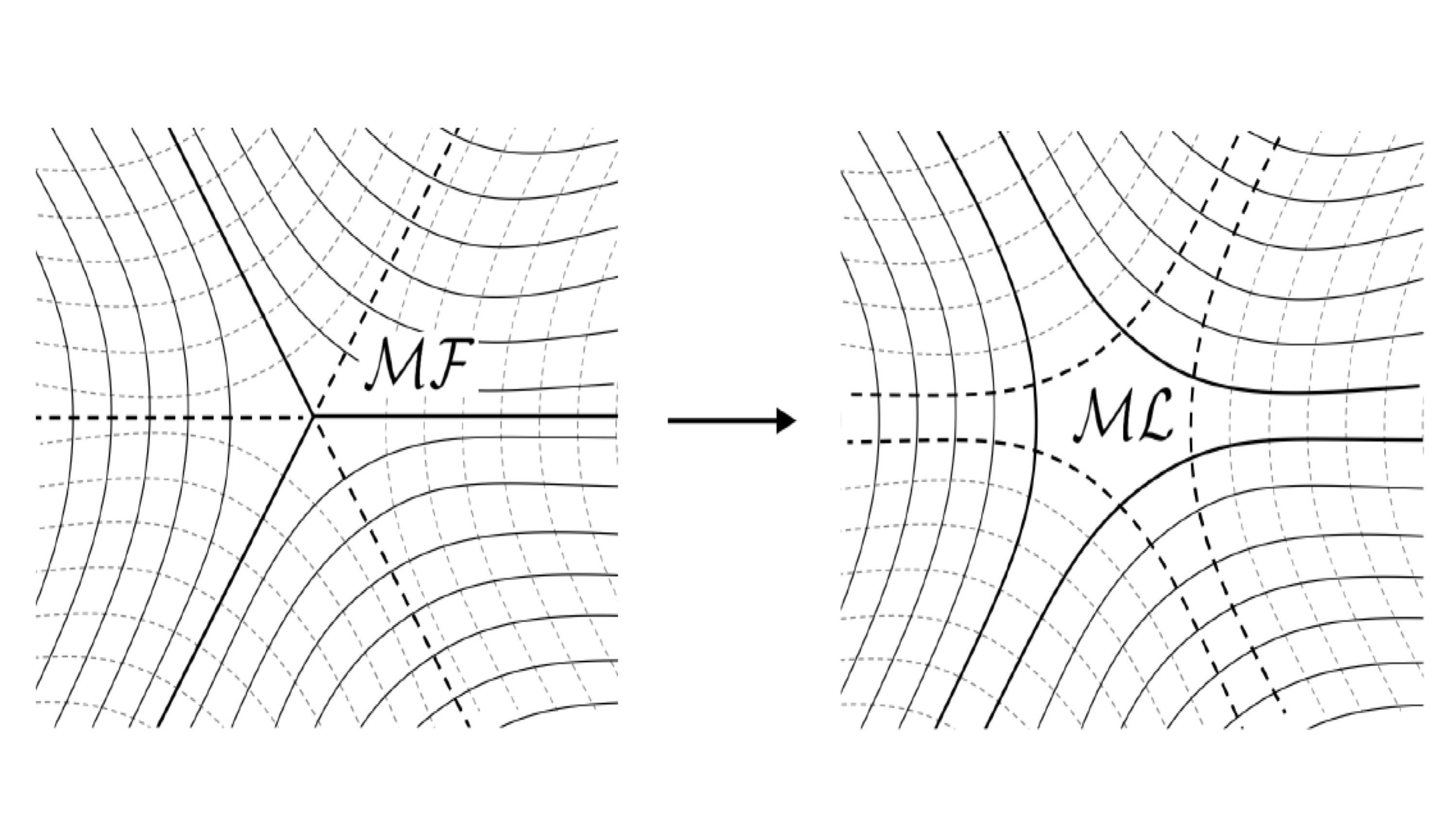}
\end{center}
\begin{center}
\caption{The Straightening map which transforms a singular measured foliation (and the transverse one) to a measured lamination (and respective transverse one).}
\label{fig:foli}
\end{center}
\end{figure}
\end{center}

\noindent Let us first define a holomorphic quadratic differential on a Riemann surface $\Sigma_{p}$. A holomorphic quadratic differential is a holomorphic section of the symmetric square of the holomorphic cotangent bundle of $\Sigma_{p}$. It may be defined locally as follows. Let $\{z_{a}:U_{a}\to \mathbb{C}\}$ be an atlas for $\Sigma_{p}$. A holomorphic quadratic differential $\Phi$ on $\Sigma_{p}$ is locally expressible on the chart $z_{a}$ as $\Phi_{a}(z_{a})dz^{2}_{a}$ with the following properties: [1] $\Phi_{a}: z_{a}(U_{a})\to \mathbb{C}$ is holomorphic, i.e., $\frac{\partial\Phi_{a}}{\partial\bar{z}_{a}}=0$, and [2] $\Phi_{a}(z_{a})(\frac{dz_{a}}{dz_{b}})^{2}=\Phi_{b}(z_{b})$ for two different overlapping charts $z_{a}:U_{a}\to\mathbb{C}$ and $z_{b}:U_{b}\to\mathbb{C}$. The second condition precisely states the invariance of $\Phi dz^{2}$ under coordinate transformations. Let us denote the space of holomorphic quadratic differentials on $\Sigma_{p}$ by $\mathcal{QD}$. By the famous theorem of Hubbard and Masur \cite{HM79}, there is a homeomorphism between the space of holomorphic quadratic differential $\mathcal{QD}$ and the space of measured foliations $\mathcal{MF}$ on $\Sigma_{p}$. One may simply associate a vertical or horizontal foliation with $\Phi\in \mathcal{QD}$ (up to isotopy and Whitehead moves; see \cite{wolf1989teichmuller} for details about Whitehead moves). For details, the reader is referred to \cite{FM11}. For now, we will only need this homeomorphism property. Given a holomorphic quadratic differential $\Phi(z) dz^{2}$ in some chart, the transverse measures of a non-trivial element $\mathcal{A}$ of $\pi_{1}(\Sigma_{p})$ (except at the zeros of $\Phi$, which correspond to the singularities of the foliation) with respect to the vertical foliation and the horizontal foliation associated with $\Phi$ are defined as follows 
\begin{eqnarray}
\label{eq:VM}
\mu_{vert}(\mathcal{A}):=\int_{\mathcal{A}}|\mathcal{R}\left(\sqrt{\Phi(z)}dz\right)|,\\
\label{eq:HM}
\mu_{hor}(\mathcal{A}):=\int_{\mathcal{A}}|\mathcal{I}\left(\sqrt{\Phi(z)}dz\right)|,
\end{eqnarray}
where $\mathcal{R}$ and $\mathcal{I}$ denote the real and imaginary parts, respectively. We will use these definitions later while considering the Einstein flow on $\Sigma_{p}$ exclusively. Given a measured foliation, one may obtain a measured lamination via a suitable straightening map \cite{minsky1992harmonic, dumas07} (or collapsing a lamination yields a foliation). Therefore, there is a homeomorphism between $\mathcal{MF}$ and $\mathcal{ML}$. Figure (\ref{fig:foli}) depicts the mechanism of yielding a lamination from a foliation. For my purposes, we will only use the homeomorphism between $\mathcal{QD}$ and $\mathcal{MF}$. All of these spaces remain homeomorphic to each other at the level of projective spaces.      

\subsection{Harmonic Maps}
\label{harmonic}
\noindent Let us now introduce another essential ingredient of my analysis: the harmonic maps. These will be crucial later in studying the Einsteinian dynamics. Let us consider a map $\mathcal{E}: (M,g) \to (N,\rho)$ (where M and N are considered to be two closed Riemann surfaces) and define the Dirichlet energy 
\begin{eqnarray}
E[\mathcal{E};g,\rho]&=&\frac{1}{2}\int_{M}\rho_{\alpha\beta}\frac{\partial \mathcal{E}^{\alpha}}{\partial x^{i}}\frac{\partial \mathcal{E}^{\beta}}{\partial x^{j}}g^{ij}\mu_{g}.
\end{eqnarray}
From the expression of the Dirichlet energy, it is obvious that it only depends on the conformal structure of the domain, that is, a conformal transformation $g_{ij}\mapsto e^{2\delta}g_{ij}, \delta: M\to \mathbb{R}$ leaves $E$ invariant. Harmonic maps are defined to be the critical points of this Dirichlet energy functional in the space of $\mathcal{E}$. The critical points of $E$ may be computed as follows. On the bundle $T^{*}M\otimes \mathcal{E}^{-1}_{*}TN$ (while restricted to the image), one has the following connection
\begin{eqnarray}
\nabla_{i}A^{\alpha}_{j}:=\partial_{i}A^{\alpha}_{j}+^{N}\Gamma^{\alpha}_{\beta\gamma}A^{\beta}_{j}\frac{\partial \xi^{\gamma}}{\partial x^{i}}-^{M}\Gamma^{k}_{ij}A^{\beta}_{k},
\end{eqnarray}
for $A\in$ \{sections$(T^{*}M\otimes \mathcal{E}^{-1}_{*}TN)\}$.
Using this definition of the connection, a few lines of calculation yields the harmonicity condition
\begin{eqnarray}
g^{ij}\partial_{i}\partial_{j}\xi^{\alpha}-g^{ij}~^{M}\Gamma^{k}_{ij}\partial_{k}\xi^{\alpha}+^{N}\Gamma^{\alpha}_{\beta\gamma}\partial_{i} \xi^{\beta}\partial_{j} \xi^{\gamma}g^{ij}&=&0.
\end{eqnarray} 
From \cite{yau78, yau97}, we know that there is a harmonic map homotopic to the identity i.e., $\mathcal{E}\in\mathcal{D}_{0}$ and in fact such a map is an orientation preserving diffeomorphism. If we take $\mathcal{E}$ to be the identity map, then the harmonicity condition reduces to the following 
\begin{eqnarray}
-g^{ij}\left(\Gamma[g]^{\alpha}_{ij}-\Gamma[\rho]^{\alpha}_{ij}\right)=0.
\end{eqnarray}
This condition will be of extreme importance when we fix the spatial gauge of the Einstein equations and also in the later part of the analysis. The Dirichlet energy of this identity map is computed to be 
\begin{eqnarray}
E[id;g,\rho]&=&\frac{1}{2}\int_{\Sigma_{p}}\rho_{ij}g^{ij}\mu_{g}.
\end{eqnarray}
Note that the conformal and diffeomorphism invariance of $E[id;g,\rho]$ allow it to be a function on the Teichm\"uller space of $\Sigma_{p}$ and in particular a proper function (i.e., the inverse images of the compact sets are compact)\cite{tromba2012teichmuller,yau97}.

\section{Einstein flow on $\Sigma_{p}\times \mathbb{R}$}
\noindent We will use the ADM formalism of general relativity in order to obtain a Cauchy problem for `$2+1$' gravity. The ADM formalism of `2+1' gravity splits the spacetime described by a `2+1' dimensional Lorentzian manifold $(\widetilde{M},\widetilde{g})$ into $\mathbb{R}\times \Sigma_{p}$ with each level set $\{t\}\times \Sigma_{p}$ of the time function $t$ being an orientable 2-manifold diffeomorphic to a Cauchy hypersurface (assuming the spacetime admits a Cauchy hypersurface) and equipped with a Riemannian metric induced by an embedding $i_{t}:\Sigma_{p}\hookrightarrow \widetilde{M}$. Such a split may be executed by introducing a lapse function $N$ and shift vector field $X$ belonging to suitable function spaces and defined such that
\begin{eqnarray}
\partial_{t}&=&N\widehat{n}+X
\end{eqnarray}
with $t$ and $\widehat{n}$ being time and a hypersurface orthogonal future-directed timelike unit vector i.e., $\widetilde{g}(\widehat{n},\widehat{n})=-1$, respectively. The above splitting writes the spacetime metric $\widetilde{g}$ in local coordinates $\{x^{\alpha}\}_{\alpha=0}^{2}=\{t,x^{1},x^{2}\}$ as 
\begin{eqnarray}
\label{eq:spacetime}
\widetilde{g}&=&-N^{2}dt\otimes dt+g_{ij}(dx^{i}+X^{i}dt)\otimes(dx^{j}+X^{j}dt)
\end{eqnarray} 
where $g_{ij}dx^{i}\otimes dx^{j}$ is the induced Riemannian metric on $\Sigma_{p}$. In order to describe the embedding of the Cauchy hypersurface $\Sigma_{p}$ into the spacetime $\widetilde{M}$, one needs information about how the hypersurface is curved in the ambient spacetime. Thus, one needs the second fundamental form $k$ defined as 
\begin{eqnarray}
K_{ij}&=-\frac{1}{2N}(\partial_{t}g_{ij}-(L_{X}g)_{ij}),
\end{eqnarray} 
the trace of which ($tr_{g}K=\tau=g^{ij}K_{ij}$, $g^{ij}\frac{\partial}{\partial x^{i}}\otimes \frac{\partial}{\partial x^{j}}:=g^{-1}$) is the mean extrinsic curvature of $\Sigma_{p}$ in $\widetilde{M}$ and $L$ denotes the Lie derivative operator.
The vacuum Einstein equations 
 \begin{equation}
 \label{eq:ein}
 R_{\mu\nu}(\widetilde{g})-\frac{1}{2}R(\widetilde{g})\widetilde{g}_{\mu\nu}=0 
 \end{equation}
may now be expressed as the evolution and constraint (Gauss and Codazzwe equations) equations of $g$ and $k$
\begin{eqnarray}
\label{eq:evol1}
\partial_{t}g_{ij}&=&-2NK_{ij}+(L_{X}g)_{ij},\\
\label{eq:evol2}
\partial_{t}K_{ij}&=&-\nabla_{i}\nabla_{j}N+N(R_{ij}+\tau K_{ij}-2k^{k}_{i}k_{jk})+(L_{X}k)_{ij},\\
\label{eq:HC}
0&=&R(g)-|K|^{2}+(tr_{g}K)^{2},\\
\label{eq:momentum}
0&=&\nabla^{i}K_{ij}-\nabla_{j}tr_{g}K.
\label{eq:cons}
\end{eqnarray}
Note that there is no canonical way to split the spacetimes, that is, the choice of a spacelike hypersurface is not unique. In order to choose a slice and study its evolution under the Einstein flow, we must fix the gauge. In my case, the most convenient choice is the constant mean extrinsic curvature spatial harmonic gauge used by \cite{moncrief03}. In this gauge, $\tau=tr_{g}K$ is constant throughout the hypersurface ($\partial_{i}\tau=0$) and therefore is chosen to play the role of time
\begin{eqnarray}
t=monotonic~function~of~\tau.
\end{eqnarray}
Spatial harmonic gauge is precisely the vanishing of the tension vector field $-g^{ij}\left(\Gamma[g]^{k}_{ij}-\Gamma[\hat{g}]^{k}_{ij}\right)$, where $\hat{g}$ is an arbitrary background metric or in other words, the harmonicity of the identity map defined in the previous section. This choice of gauge yields the following two elliptic equations for the lapse function and the shift vector field, respectively
\begin{eqnarray}
\label{eq:lapse}
\Delta_{g}N+N(|K^{TT}|^{2}_{g}+\frac{\tau^{2}}{2})=\partial_{t}\tau,\\
\label{eq:shift}
\Delta_{g}X^{i}-R^{i}_{j}X^{j}=(\nabla^{i}N)\tau-2\nabla^{j}NK^{i}_{j}+(2NK^{jk}-2\nabla^{j}X^{k})\\\nonumber (\Gamma[g]^{i}_{jk}-\hat{\Gamma}[\hat{g}]^{i}_{jk}).
\end{eqnarray}
This Cauchy problem (with initial data $(g_{0},k_{0})$) with constant mean extrinsic curvature and spatially harmonic gauge is referred to as ‘CMCSH Cauchy’ problem.
\subsection{\textbf{Well-posedness:}}
\noindent \cite{moncrief03} proved a local well-posedness theorem for the Cauchy problem for a family of elliptic- hyperbolic systems that included the `$n+1$' dimensional vacuum Einstein equations in CMCSH gauge, $n\geq2$. They also proved the conservation of gauges and constraints. In addition to the local well-posedness, \cite{anderson1997global} proved a global existence theorem for the expanding solutions in the same gauge by controlling the Dirichlet energy of an associated harmonic map for any $\tau\in(-\infty,0)$. Therefore, the well-posedness of the Cauchy problem is established and we do not wish to repeat the same here. Interested readers are referred to these articles.    

\subsection{\textbf{Reduced Dynamics}}
\label{reduced}
\noindent Given a scalar function $\varphi: \Sigma_{p}\to \mathbb{R}$, we define a set of conformal variables $(\gamma,k^{TT})$ ($k^{TT}$ is transverse-traceless with respect to the metric $\gamma$) in terms of the physical variables $(g,k^{TT}$) by setting 
\begin{eqnarray}
\label{eq:conformal}
(g_{ij},K^{TTij})&=&(e^{2\varphi}\gamma_{ij}, e^{-4\varphi}k^{TTij}),
\end{eqnarray}
where $R(\gamma)=-1$ (such a unique $\gamma$ exists if genus$(\Sigma_{p})>1)$) and the second fundamental form is written as follows 
\begin{eqnarray}
K=K^{TT}+\frac{\tau}{2}g,
\end{eqnarray}
by using the momentum constraint with $K^{TT}$ being transverse-traceless with respect to $g$. Here $k^{TT}$ is transverse-traceless with respect to $\gamma$, that is,
\begin{eqnarray}
\nabla[\gamma]_{j}k^{TTij}&=&0,\\
\gamma_{ij}k^{TTij}&=&0,
\end{eqnarray}
if and only if $K^{TT}$ is transverse-traceless with respect to $g$. Naturally 
\begin{eqnarray}
k^{TT}_{ij}=K^{TT}_{ij},\\
k^{TTij}=\gamma^{ik}\gamma^{jl}k^{TT}_{kl}.
\end{eqnarray}
$\varphi$ can be found by solving the Hamiltonian constraint which now takes the form of the following semilinear elliptic PDE namely the  Lichnerowicz equation 
\begin{eqnarray}
\label{eq:constraint}
-2\Delta_{\gamma}\varphi+1+e^{-2\varphi}|k^{TT}|^{2}_{\gamma}-\frac{e^{2\varphi}\tau^{2}}{2}&=&0.
\end{eqnarray}
Using the sub and super solution technique \cite{choquet2009general, Oxford University Press}, it is established that there is a unique solution $\varphi[\gamma, k^{TT},\tau]$ of the Lichnerowicz equation. Indeed, this equation will be crucial to my analysis of proving the main theorem. The phase space of the reduced dynamics now may be defined as $\{(\gamma_{ij},k^{TTij})| \gamma\in \mathcal{M}_{-1},  tr_{\gamma}k^{TT}=0=\nabla[\gamma]_{j}k^{TTij}\}$. In reality, the true dynamics assumes a metric lying in the orbit space $\mathcal{M}_{-1}/D_{0}$, $D_{0}$ being the group of diffeomorphisms (of $\Sigma_{p}$) isotopic to identity. This is a consequence of the fact that if $\gamma_{ij}\in\mathcal{M}_{-1}, k^{TTij}, \varphi, N$, and $X^{i}$ solve the Einstein equations, so do $((\phi^{-1})^{*}\gamma)_{ij}, (\phi_{*}k^{TT})^{ij}, (\phi^{-1})^{*}\varphi=\varphi\circ \phi^{-1}, (\phi^{-1})^{*}N=N\circ \phi^{-1}$, and $(\phi_{*}X)^{i}$, where $\phi\in D_{0}$ and $^{*}$, and $_{*}$ denote the pullback and push-forward operations (time-independent) on the cotangent and tangent bundles of $M$, respectively. Let us now consider a time-dependent $\phi_{t}\in \mathcal{D}_{0}$ and go back to the un-scaled dynamical equation (\ref{eq:evol1}) (note that the un-scaled fields $(g,K,N,X)$) solve the Einstein's dynamical and constraint equations (\ref{eq:evol1}-\ref{eq:cons}) iff $(\gamma,
k,\varphi,N,X)$ solve the reduced equations)
\begin{eqnarray}
\partial_{t}((\phi^{-1}_{t})^{*}g)_{ij}=-2(\phi^{-1}_{t})^{*})(NK_{ij})+(L_{\phi_{t*}X}(\phi^{-1}_{t})^{*}g)_{ij},\\\nonumber
(\phi^{-1}_{t})^{*}\partial_{t}g_{ij}+(\partial_{t}(\phi^{-1}_{t})^{*})g_{ij}=-2(\phi^{-1}_{t})^{*}(NK_{ij})+\frac{\partial}{\partial s}((\phi^{-1}_{t}\varphi^{X}_{s}\phi_{t})^{*}(\phi^{-1}_{t})^{*}g)|_{s=0},\\\nonumber
(\phi^{-1}_{t})^{*}\partial_{t}g_{ij}+(\partial_{s}(\phi^{-1}_{t+s})^{*})g_{ij}|_{s=0}=-2(\phi^{-1}_{t})^{*}(NK_{ij})+(\phi^{-1}_{t})^{*}(L_{X}g)_{ij},\\\nonumber
(\phi^{-1}_{t})^{*}\partial_{t}g_{ij}+(\phi^{-1}_{t})^{*}(L_{Y}g)_{ij}=-2(\phi^{-1}_{t})^{*}(NK_{ij})+(\phi^{-1}_{t})^{*}(L_{X}g)_{ij},\\\nonumber
(\phi^{-1}_{t})^{*}\left\{\partial_{t}g_{ij}=-2NK_{ij}+(L_{X-Y}g)_{ij}\right\}.
\end{eqnarray}
Here $Y$ is the vector field associated with the flow $\phi_{t}$ and $\varphi^{X}_{s}$ is the flow of the shift vector field $X$.
A similar calculation for the evolution equation for the second fundamental form shows that if we make a transformation $X\mapsto X+Y$, the Einstein evolution, and constraint (due to their natural spatial covariance nature) equations are satisfied by the transformed fields. The action of $\phi_{t}$ on the un-scaled fields naturally extends to the conformally scaled fields. Therefore, the true reduced dynamics occur on the quotient space $\mathcal{M}_{-1}/\mathcal{D}_{0}$. Now, $\mathcal{M}_{-1}/D_{0}$ is precisely the Teichm\"uller space of $\Sigma_{p}$ and following \cite{tromba2012teichmuller}, the transverse-traceless tensor $k^{TT}$ models the tangent space at $\gamma$. Therefore, we obtain the Teichm\"uller space ($6p-6$ dimensional) $\mathcal{T}\Sigma_{p}$ as the configuration space, while the cotangent bundle ($12p-12$ dimensional) of $\mathcal{T}\Sigma_{p}$ serves as the phase space of the reduced dynamics. This is precisely what was stated previously in section 2 while relating the full solution space of the vacuum Einstein equations and the Teichm\"uller space through its algebraic topologic definition.    

\noindent Now we will obtain a series of estimates which will be useful for the later analysis. The following lemma provides a point-wise estimate for the conformal factor $e^{2\varphi}$.\\
\begin{lemma}
\label{scalarestimate} Let $\varphi: \Sigma_{p}\to \mathbb{R}$ solves the Lichnerowicz equation (\ref{eq:constraint}). Then $e^{2\varphi}:=e^{2\varphi(\tau,k^{TT},\gamma)}$ verifies the following point-wise estimate 
\begin{eqnarray}
\label{eq:estimate1}
\frac{2}{\tau^{2}}\leq e^{2\varphi}\leq \frac{1+\sqrt{1+2\tau^{2}\sup_{\Sigma_{p}}|k^{TT}|^{2}_{\gamma}(\tau)}}{\tau^{2}}~~~~\forall~\tau\in(-\infty,0).
\end{eqnarray}
\end{lemma}
\textbf{Proof:} A standard maximum principle argument while applied to the Lichnerowicz equation (\ref{eq:constraint}) yields the following 
\begin{eqnarray}
\tau^{2}e^{4\varphi}-2e^{2\varphi}-2\sup_{\Sigma_{p}}|k^{TT}|^{2}_{\gamma}(\tau)\leq 0
\end{eqnarray}
that implies
\begin{eqnarray}
\left(e^{2\varphi}-\frac{1+\sqrt{1+2\tau^{2}\sup_{\Sigma_{p}}|k^{TT}|^{2}_{\gamma}(\tau)}}{\tau^{2}}\right)
\left(e^{2\varphi}-\frac{1-\sqrt{1+2\tau^{2}\sup_{\Sigma_{p}}|k^{TT}|^{2}_{\gamma}(\tau)}}{\tau^{2}}\right)\leq 0.
\end{eqnarray}
But, $e^{2\varphi}>0$ and therefore, we must have 
\begin{eqnarray}
e^{2\varphi}\leq \frac{1+\sqrt{1+2\tau^{2}\sup_{\Sigma_{p}}|k^{TT}|^{2}_{\gamma}(\tau)}}{\tau^{2}}.
\end{eqnarray}
Similarly, at a minimum, the following holds 
\begin{eqnarray}
\tau^{2}e^{4\varphi}-2e^{2\varphi}\geq 0,
\end{eqnarray}
that is,
\begin{eqnarray}
e^{2\varphi}\geq \frac{2}{\tau^{2}},
\end{eqnarray}
where the equality holds if and only if 
\begin{eqnarray}
k^{TT}\equiv0.
\end{eqnarray}
In summary, we have the following estimate of the conformal factor from the Lichnerowicz equation
\begin{eqnarray}
\label{eq:nonlinearelliptic}
\frac{2}{\tau^{2}}\leq e^{2\varphi}\leq \frac{1+\sqrt{1+2\tau^{2}\sup_{\Sigma_{p}}|k^{TT}|^{2}_{\gamma}(\tau)}}{\tau^{2}},
\end{eqnarray}
which will be useful later. This concludes the proof of the lemma.~~~~~~~~~~~$\Box$

\noindent Now we will obtain an estimate for  $|K^{TT}|^{2}_{g}=e^{-4\varphi}|k^{TT}|^{2}_{\gamma}$. In order to do so, we first obtain an elliptic equation for $|K^{TT}|^{2}_{g}$. The following lemma provides the necessary elliptic equation $|K^{TT}|^{2}_{g}$.

\noindent \begin{proposition} Let $K:=K^{TT}+\frac{1}{2}\tr_{g}K g$ be a solution of the momentum constraint equation (\ref{eq:momentum}) in CMC gauge $\partial_{i}\tau:=\partial_{i}\tr_{g}K=0$, then $|K^{TT}|^{2}_{g}$ satisfies the following quasi-linear elliptic equation on a constant time hypersurface
\begin{eqnarray}
-\Delta_{g}(|K^{TT}|^{2}_{g})-2|K^{TT}|^{2}_{g}(|K^{TT}|^{2}_{g}-\frac{1}{2}\tau^{2})
&=&2\nabla[g]_{k}(K^{TT}_{ij})\nabla[g]^{k}(K^{TTij}).
\end{eqnarray}
\end{proposition}
\textbf{Proof:}
Note that in 2 dimensions, the momentum constraint 
\begin{eqnarray}
\nabla[g]_{j}K^{j}_{i}-\nabla_{i}tr_{g}K=0
\end{eqnarray}
implies that $K$ is a Codazzi tensor \cite{moncrief2007relativistic, anderson1997global} i.e.,
\begin{equation}
\nabla[g]_{j}K^{i}_{k}-\nabla[g]_{k}K^{i}_{j}=0.    
\end{equation}
After substituting the decomposition $K=K^{TT}+\frac{\tr_{g}K}{2}g$ in the Codazzi equation, Covariant divergence yields 
\begin{eqnarray}
\nabla[g]^{j}\nabla[g]_{j}K^{TTi}_{k}-\nabla[g]^{j}\nabla[g]_{k}K^{TTi}_{j}=0,\\\nonumber 
\nabla[g]^{j}\nabla[g]_{j}K^{TTi}_{k}-\nabla[g]_{k}\nabla[g]_{j}K^{TTij}-R[g]^{i}~_{mjk}K^{TTmj}-R[g]^{j}~_{mjk}K^{TTim}=0,
\end{eqnarray}
which upon utilizing $\nabla[g]_{j}K^{TTij}=0$ and $R[g]^{i}~_{mjk}=\frac{R(g)}{2}(\delta^{i}_{j}g_{mk}-\delta^{i}_{k}g_{mj})$ reduces to 
\begin{eqnarray}
\nabla[g]^{j}\nabla[g]_{j}K^{TTi}_{k}=R(g)K^{TTi}_{k}.
\end{eqnarray}
$\Delta_{g}(|K^{TT}|^{2}_{g})$ may be evaluated as follows 
\begin{eqnarray}
\Delta_{g}(|K^{TT}|^{2}_{g})=-\nabla[g]^{j}\nabla[g]_{j}|K^{TT}|^{2}_{g}=-\nabla[g]^{j}\nabla[g]_{j}(K^{TTi}_{k}K^{TTk}_{i})\\\nonumber 
=-2(\nabla[g]^{j}\nabla[g]_{j}K^{TTi}_{k})K^{TTk}_{i}-2\nabla[g]^{j}K^{TTi}_{k}\nabla[g]_{j}K^{TTk}_{i},\\\nonumber
=-2R(g)|K^{TT}|^{2}_{g}-2\nabla[g]^{j}K^{TTi}_{k}\nabla[g]_{j}K^{TTk}_{i},\\\nonumber
=-2(|K^{TT}|^{2}_{g}-\frac{\tau^{2}}{2})|K^{TT}|^{2}_{g}-2\nabla[g]^{j}K^{TTi}_{k}\nabla[g]_{j}K^{TTk}_{i},
\end{eqnarray}
i.e.,
\begin{eqnarray}
-\Delta_{g}(|K^{TT}|^{2}_{g})-2|K^{TT}|^{2}_{g}(|K^{TT}|^{2}_{g}-\frac{1}{2}\tau^{2})
&=&2\nabla[g]_{k}(K^{TT}_{ij})\nabla[g]^{k}(K^{TTij}).
\end{eqnarray}
Here, we have used the Hamiltonian constraint (\ref{eq:HC}) $|K^{TT}|^{2}_{g}=R(g)+\frac{\tau^{2}}{2}$. This concludes the proof of the lemma. ~~~~~~~~~~~~~$\Box$

\noindent Remarkably, the quasi-linear term appearing in the right-hand side of the elliptic equation (\ref{eq:nonlinearelliptic}) has a favorable sign that is conducive to an application of a standard maximum principle.\\
\begin{lemma} 
\label{momentumestimate} \textit{$|K^{TT}|^{2}_{g}$ satisfies the estimate 
\begin{eqnarray}
|K^{TT}|^{2}_{g}\leq \frac{\tau^{2}}{2}
\end{eqnarray}
for all $\tau\in (-\infty,0)$.
}
\end{lemma}
\textbf{Proof:}
The quasi-linear term satisfies $\nabla[g]_{k}(K^{TT}_{ij})\nabla[g]^{k}(K^{TTij})\geq0$. Application of a standard maximum principle argument yields 
\begin{eqnarray}
\label{eq:estimate2}
|K^{TT}|^{2}_{g}\leq \frac{\tau^{2}}{2}.~~~~~~~\Box
\end{eqnarray}

\noindent Lastly, we will obtain an estimate for the lapse function after choosing the following time coordinate
\begin{eqnarray}
\label{eq:time}
t:=-\frac{1}{\tau}.
\end{eqnarray}
The allowed time range in this coordinate is $(0,\infty)$. The lapse equation (\ref{eq:lapse}) now reads 
\begin{eqnarray}
\label{eq:lapseproxy}
\Delta_{g}N+N(|K^{TT}|^{2}_{g}+\frac{\tau^{2}}{2})=\tau^{2}.
\end{eqnarray}

\begin{corollary}
\label{lapseestimate}
The Lapse function $N$ verifies the estimate $1\leq N\leq 2$
\end{corollary}
\begin{proof}
A standard maximum principle argument applied to the lapse equation (\ref{eq:lapseproxy}) together with the estimate (\ref{eq:estimate2}) yields the desired estimate.
\end{proof}

Now we will describe Moncrief's ray structure \cite{moncrief2007relativistic} of the Teichm\"uller space, which will be of crucial in obtaining the main result. The ray structure defined by Moncrief is the following equation 
\begin{eqnarray}
\label{eq:gauss}
\rho_{ij}&=&|K|^{2}_{g}g_{ij}+2\tau(K_{ij}-\frac{1}{2}\tau g_{ij})\\\nonumber
&=&(|K^{TT}|^{2}_{g}+\frac{\tau^{2}}{2})g_{ij}+2\tau K^{TT}_{ij}\\\nonumber
&=&(e^{-4\varphi}|k^{TT}|^{2}_{\gamma}+\frac{\tau^{2}}{2})e^{2\varphi}\gamma_{ij}+2\tau k^{TT}_{ij}
\end{eqnarray}  
together with an associated Hamilton-Jacobi equation. Here $\rho$ is a fixed metric satisfying $R(\rho)=-1$ (and therefore lies inside the Teichm\"uller space) and $g_{ij}$ is solved in terms of $\rho_{ij}$. This computes the end point of a ray in terms of the data along the ray. For the detailed derivation of this expression, one may consult the relevant section of \cite{moncrief2007relativistic}. This is designated in \cite{moncrief2007relativistic} as the `\textbf{Gauss}' map equation. For my purpose, the derivation of this map is tangential and hence, we do not wish to repeat the same here. The vital question is whether such $(g_{ij},K^{TTij},N,X)$ actually solves the Einstein equations for all $\tau$ given an initial $(g_{ij0},K^{TTij}_{0},N_{0},X_{0})$ satisfying the constraint equations. This is equivalent to solving for conformal variables $(\gamma_{ij},k^{TTij}, \varphi)$ and associated lapse function $N$ and shift vector field $X$. This is exactly shown in \cite{moncrief2007relativistic} through studying the associated Hamilton Jacobi equation for the reduced dynamics. When this lagrangian formulation is cast into a more natural Hamiltonian one, one clearly sees that the original Einstein-Hilbert action may be written as follows 
\begin{eqnarray}
S&=&\int_{I\subset \mathbb{R}}\int_{\Sigma_{p}}\left(\mu_{g}(-K^{ij}+\tau g^{ij})\frac{\partial g_{ij}}{\partial t}-N\mathcal{H}-X^{i}\mathcal{P}_{i}\right)d^{2}xdt,
\end{eqnarray}
where $\mathcal{H}:=\mu_{g}K^{TT}_{ij}K^{TTij}-\frac{1}{2}\tau^{2}\mu_{g}-\mu_{g}R(g)$, and $\mathcal{P}_{i}:=2\nabla[g]_{j}(\mu_{g}K^{j}_{i}-\tau\mu_{g}\delta^{j}_{i})$. Note that vanishing of $\mathcal{H}$ and $\mathcal{P}_{i}$ is precisely equivalent to $(g_{ij},K^{ij})$ satisfying the Hamiltonian and momentum constraints. When both of these constraints are satisfied we obtain the reduced action 
\begin{eqnarray}
S_{reduced}&=&\int_{I\subset \mathbb{R}}\int_{\Sigma_{p}}\mu_{g}(-K^{ij}+\tau g^{ij})\frac{\partial g_{ij}}{\partial t}d^{2}xdt,
\end{eqnarray}
which through the conformal transformation (\ref{eq:conformal}) becomes 
\begin{eqnarray}
\label{eq:reduced}
S_{reduced}&=&\int_{I\subset \mathbb{R}}\left(\int_{\Sigma_{p}}(-\mu_{\gamma}k^{TTij}\frac{\partial \gamma_{ij}}{\partial t}-\frac{\partial\tau}{\partial t}\mu_{g})d^{2}x\right)dt,
\end{eqnarray}
where the boundary terms (`in time') are ignored, because, they do not contribute to the equations of motions at the classical level. The Hamiltonian of this reduced dynamics can be read off as follows from the expression of the previous action
\begin{eqnarray}
H_{reduced}&=&\int_{\Sigma_{p}}\frac{\partial \tau}{\partial t}\mu_{g}.
\end{eqnarray}
Substituting the time coordinate from equation (\ref{eq:time}) into the expression of the reduced Hamiltonian together with the Hamiltonian constraint yields 
\begin{eqnarray}
H_{reduced}&=&2\int_{\Sigma_{p}}|K^{TT}|^{2}_{g}-8\pi\chi,
\end{eqnarray}
where $\chi=2(1-g)<0$ is the Euler characteristics of $\Sigma_{p}$. This reduced Hamiltonian can be related to the Dirichlet energy of the Gauss map. The Dirichlet energy (conformally invariant on the domain) associated to the Gauss map (\ref{eq:gauss}) is given as
\begin{eqnarray}
\label{eq:energy}
E[id;g,\rho]&=&\frac{1}{2}\int_{\Sigma}\mu_{g}g^{ij}\rho_{ij}=\frac{1}{2}\int_{\Sigma}\mu_{\gamma}\gamma^{ij}\rho_{ij}=E[id;\gamma,\rho]\\\nonumber
&=&2\int_{\Sigma}|K^{TT}|^{2}_{g}\mu_{g}-4\pi\chi.
\end{eqnarray}
Therefore, we have the following relation between the Dirichlet energy of the Gauss map and the reduced Hamiltonian of the dynamics 
\begin{eqnarray}
\label{eq:relation}
H_{reduced}=E[id;\gamma,\rho]-4\pi\chi.
\end{eqnarray}
Let us consider that the Teichm\"uller space $\mathcal{T}\Sigma_{p}$ is parametrized by $\{q_{\alpha}\}_{\alpha=1}^{6p-6}$, which may be of the Fenchel-Neilsen type (see \cite{farb2011primer} for details about Fenchel-Neilsen parametrization).
Now we observe the following 
\begin{eqnarray}
\frac{\partial E[id;\gamma(q),\rho]}{\partial q_{\alpha}}&=&\frac{1}{4}\int_{\Sigma_{p}}\mu_{\gamma}\left(\gamma^{mn}\gamma^{ij}\rho_{ij}-2\gamma^{im}\gamma^{jn}\rho_{ij}\right)\frac{\partial \gamma_{mn}}{\partial q_{\alpha}},
\end{eqnarray}
which after substituting $\rho_{ij}=(|K^{TT}|^{2}_{g}+\frac{\tau^{2}}{2})g_{ij}+2\tau K^{TT}_{ij}=(e^{-4\varphi}|k^{TT}|^{2}_{\gamma}+\frac{\tau^{2}}{2})e^{2\varphi}\gamma_{ij}+2\tau k^{TT}_{ij}$ yields 
\begin{eqnarray}
\label{eq:relation1}
\frac{\partial E[id;\gamma(q),\rho]}{\partial q_{\alpha}}&=&-\tau\int_{\Sigma_{p}}\mu_{\gamma}k^{TTij}\frac{\partial \gamma^{mn}}{\partial q_{\alpha}}.
\end{eqnarray}
Now let us go back to equation (\ref{eq:reduced}) and substitute $\gamma=\gamma(q)$. We immediately obtain 
\begin{eqnarray}
S_{reduced}&=&\int_{I\subset \mathbb{R}}\left((\int_{\Sigma_{p}}-\mu_{\gamma}k^{TTij}\frac{\partial \gamma_{ij}}{\partial q_{\alpha}})\dot{q}_{\alpha}-H_{reduced}(\gamma(q),p,\rho)\right)dt\\\nonumber
&=&\int_{I\subset \mathbb{R}}\left(p^{\alpha}\dot{q}_{\alpha}-H_{reduced}(\gamma(q),p,\rho)\right)dt,
\end{eqnarray}
which upon utilizing equations (\ref{eq:relation}) and (\ref{eq:relation1}) leads to 
\begin{eqnarray}
\frac{\partial H_{reduced}(\gamma(q),p,\rho)}{\partial q_{\alpha}}&=&\tau p^{\alpha}.
\end{eqnarray}
Here $\{(q_{\alpha},p^{\alpha})\}_{\alpha=1}^{6p-6}$ parametrizes the phase space i.e., the co-tangent bundle of $\mathcal{T}\Sigma_{p}$. Now using the time defined in $(\ref{eq:time})$, we may construct a principle functional after substituting $T=-\frac{1}{\tau}$ 
\begin{eqnarray}
\mathcal{S}(q,\gamma(q),\rho)&=&-T(E[id;\gamma(q),\rho]-4\pi\chi)
\end{eqnarray}
which then clearly satisfies
\begin{eqnarray}
p^{\alpha}&=&\frac{\partial \mathcal{S}}{\partial q_{\alpha}},\\
-\frac{\partial \mathcal{S}}{\partial T}&=&E[id;\gamma(q),\rho]-4\pi\chi=H_{reduced}(q,p,\gamma(q)),
\end{eqnarray}
that is, $\mathcal{S}$ satisfies the Hamilton-Jacobi equation 
\begin{eqnarray}
\label{eq:HJ}
-\frac{\partial \mathcal{S}}{\partial T}&=&H_{reduced}(q,p,\gamma(q))
\end{eqnarray}
for all $T\in(0,\infty)$. In other words, $\mathcal{S}$ is dynamically complete. For detailed analysis (arguments underlying dynamical completeness of $\mathcal{S}$), the reader is referred to the relevant sections of \cite{moncrief2007relativistic}. Here we only require the fact that through the solution of this Hamilton-Jacobi equation, the Gauss map equation defined in (\ref{eq:gauss}) solves the Einstein equation for all $T\in (0,\infty)$ or equivalently for all $\tau\in(-\infty,0)$ and defines a ray-structure based at $\rho$ of the Teichm\"uller space parametrized by the transverse-traceless conformally invariant 2-tensor $k^{TT}_{ij}$. 

\subsection{Proof of the Theorem 1}
\noindent  {An implicit solution \cite{moncrief2007relativistic} of the Gauss map equation (\ref{eq:gauss}) gives 
\begin{eqnarray}
\gamma^{ij}=e^{2\varphi}g^{ij}=e^{2\varphi}\left(\frac{2\tau^{3}}{\mu_{\rho}}\frac{\rho^{ik}\mu_{\gamma}\gamma^{jl}k^{TT}_{kl}}{1+\sqrt{1+\frac{2\tau^{2}\mu^{2}_{\gamma}|k^{TT}|^{2}_{\gamma}}{\mu_{\rho}^{2}}}}+\tau^{2}\frac{\sqrt{1+\frac{2\tau^{2}\mu^{2}_{\gamma}|k^{TT}|^{2}_{\gamma}}{\mu_{\rho}^{2}}}}{1+\sqrt{1+\frac{2\tau^{2}\mu^{2}_{\gamma}|k^{TT}|^{2}_{\gamma}}{\mu_{\rho}^{2}}}}\rho^{ij}\right)\nonumber.
\end{eqnarray}
Using this equation (which is effectively the same as the Gauss map equation), \cite{moncrief2007relativistic} constructed a fully non-linear elliptic equation of Monge-Ampere type and showed that a unique solution of such equation exists. Recently \cite{tam2019fully} showed using a direct analytic technique that such a unique solution exists for all $\tau\in (-\infty,0)$. Essentially, these analyses are in a sense complementary to the Hamilton-Jacobi theory and provide a more explicit description of the ray structure of the Teichm\"uller space. Analyzing the associated Monge-Ampere equation, \cite{moncrief2007relativistic} explicitly showed that every non-trivial solution curve of the reduced dynamics in the configuration space ($\mathcal{T}\Sigma_{p}$) approaches a point ($\rho$) lying in the interior of the Teichm\"uller space, that is, 
\begin{eqnarray}
\lim_{\tau\to 0^{-}}\gamma^{ij}&=&\rho^{ij}.
\end{eqnarray}
Note that the choice of $\rho$ is arbitrary as long as it does not leave the compact sets of $\mathcal{T}\Sigma_{p}$, and therefore, one may vary $\rho$ over $\mathcal{T}\Sigma_{p}$ to obtain the full ray-structure of the Teichm\"uller space. We do not provide the complete calculations regarding the $\tau\to 0^{-}$ behavior of the solution curve as it is derived and described in detail by Moncrief in \cite{moncrief2007relativistic}. Readers are referred to the relevant sections of the same. We only need the information that the Gauss map equation together with the Hamilton-Jacobi equation indeed describes ray structures of the Teichm\"uller space and every such ray solves the reduced Einstein equation. Forward time asymptotics of each such ray corresponds to an interior point which also realizes the infimum of the Dirichlet energy (and the reduced Hamiltonian). Each member of a family of rays which asymptotically approaches the point $\rho\in\mathcal{T}\Sigma_{p}$ corresponds to a unique choice of $k^{TT}$ and none of the two rays of the same family intersect each other (except at $\rho$, where they approach as $\tau\to0^{-}$). The following is a sketch of the argument without being completely rigorous.}  

 {First we observe the following monotonic decay of the Dirichlet energy in the time-forward direction
 \begin{eqnarray}
 \label{monotonicity}
 \frac{d}{dt}E[id;\gamma,\rho]&=&2\frac{d}{dt}\int_{\Sigma}|K^{TT}|^{2}_{g}\mu_{g},\\
 &=&2\tau\int_{\Sigma}N|K^{TT}|^{2}_{g}\mu_{g}<0\nonumber.
 \end{eqnarray}
 $\frac{d}{dt}E[id;g,\rho]\equiv0$ if and only if $K^{TT}\equiv0$ (or $k^{TT}\equiv 0$) and $\frac{d}{dt}E[id;\gamma,\rho]\to0$ at the limit $\tau\to 0$.
 Now recall the expression for the Dirichlet energy (\ref{eq:energy}) 
 \begin{eqnarray}
 \label{eq:bound}
  E[id;g,\rho]&=&\frac{1}{2}\int_{\Sigma}\mu_{g}g^{ij}\rho_{ij}=\frac{1}{2}\int_{\Sigma}\mu_{\gamma}\gamma^{ij}\rho_{ij}=E[id;\gamma,\rho]\\\nonumber
&=&2\int_{\Sigma}|K^{TT}|^{2}_{g}\mu_{g}-4\pi\chi\geq -4\pi\chi.   
 \end{eqnarray}
Firstly, consider the initial data $(g_{0},K_{0})$ (or equivalently $(\gamma_{0},k^{TT}_{0},\tau_{0},\varphi_{0})$ such that\\\nonumber $E[id;g_{0},\rho]=E[id;\gamma_{0},\rho]=E_{0}<\infty$. Along any solution ray with initial data $(\gamma_{0},k^{TT}_{0},\tau_{0},\varphi_{0})$, $E[id;\gamma,\rho]$ lies in the interval $[-4\pi\chi,E_{0}]$ due to the monotonicity property \ref{monotonicity}.  $E[id;\gamma,\rho]$ can be considered as a map from Teichm\"uller space to positive real line. Its properness implies that the inverse images of the compact sets are compact sets. Therefore, any sequence on the solution ray can not leave compact sets of the Teichm\"ller space in the forward time direction (i.e., as $\tau\to0$) due to \ref{monotonicity} and \ref{eq:bound} or more explicitly pre-image of the compact set $[-4\pi\chi,E_{0}]$ is a compact subset of the Teichm\"uller space due to the aforementioned `properness' property of the Dirichlet energy $E[id;\gamma,\rho]$. Compactness yields the extraction of a convergent subsequence from any minimizing sequence on the solution ray in the forward-in-time direction (minimizing the Dirichlet energy). One verifies that this limit satisfies Einstein's equations as follows.  
The limit of the convergent subsequence is characterized by $k^{TT}\equiv 0$ since it realizes the infimum of the Dirichlet energy. Substituting $k^{TT}=0$ into the Lichnerowicz equation (\ref{eq:constraint}) yields 
\begin{eqnarray}
-2\Delta_{\gamma}\varphi+1-\frac{e^{2\varphi}\tau^{2}}{2}&=&0,
\end{eqnarray}
which has a unique solution 
\begin{eqnarray}
e^{2\varphi}=\frac{2}{\tau^{2}}.
\end{eqnarray}
Application of a maximum principle to the Lapse equation (\ref{eq:lapse}) after substituting $k^{TT}=0$ yields $N=2$. Similarly, the shift vector field verifies $X=0$. The reduced evolution equation reads 
 \begin{eqnarray}
 \frac{\partial \gamma_{ij}}{\partial t}=e^{-2\varphi}\left(-(\partial_{t}e^{2\varphi}\gamma_{ij}-2Nk^{TT}_{ij}-e^{2\varphi}N\tau\gamma_{ij}+(L_{X}e^{2\varphi}\gamma)_{ij}\right),
 \end{eqnarray}
 which, upon substituting $e^{2\varphi}=\frac{2}{\tau^{2}}$, $k^{TT}=0, N=2$, and $X=0$ yields
 \begin{eqnarray}
 \frac{\partial \gamma_{ij}}{\partial t}=0.
 \end{eqnarray}
A few lines of simple calculation yield $\partial_{t}k^{TT}_{ij}=0$ as well. Therefore the limit solves the reduced Einstein's equations and more precisely corresponds to the fixed point solutions characterized by ($\gamma_{ij},k^{TT}_{ij}=0, N=2, X^{i}=0$), $R(\gamma)=-1$. Even though the Dirichlet energy controls the $(H^{1}(\Sigma_{p})\times L^{2}(\Sigma_{p}))$ norm of the data $(\gamma,k^{TT})$, finite dimensionality of the phase space implies that control on this norm is sufficient. 
This concludes the sketch of the first part of the proof. We do not provide a more rigorous detail due to the fact that a comprehensive detail is provided in \cite{moncrief2007relativistic}.}

 {In the second part of the theorem, we want to argue that the solution curve leaves every compact set of the Teichm\"uller space in the limit $\tau\to-\infty$. This conclusion may be obtained by studying the time evolution of the Dirichlet energy (a proper function on $\mathcal{T}\Sigma_{p}$) of the Gauss map. The time is chosen to be $t=-\frac{1}{\tau}$ (\ref{eq:time}). From equation (\ref{eq:energy}), the time derivative of the $|K^{TT}|^{2}_{g}$ reads   
\begin{eqnarray}
\frac{d}{dt}\int_{\Sigma}|K^{TT}|^{2}_{g}\mu_{g}=\frac{d}{dt}\int_{\Sigma}(\frac{\tau^{2}}{2}+R(g))\mu_{g},\\\nonumber
=\tau^{2}\frac{d}{d\tau}\int_{\Sigma}(\frac{\tau^{2}}{2}+R(g))\mu_{g},
=\tau^{3}\int_{\Sigma}\mu_{g}+\frac{\tau^{2}}{2}\int_{\Sigma}\mu_{g}(-2N\tau),\\\nonumber
=\tau\int_{\Sigma}N|K^{TT}|^{2}_{g}\mu_{g},
=-\frac{1}{t}\int_{\Sigma}N|K^{TT}|^{2}_{g}\mu_{g},\nonumber
\end{eqnarray}
where, we have used the lapse equation $\Delta_{g}N+N(|K^{TT}|^{2}_{g}+\frac{\tau^{2}}{2})=\tau^{2}$, the Hamiltonian constraint $|K^{TT}|^{2}_{g}=\frac{\tau^{2}}{2}+R(g)$, and the evolution equation $\frac{\partial g_{ij}}{\partial t}=-2NK_{ij}+(L_{X}g)_{ij}$. Utilizing the estimate of the lapse function (corollary \ref{lapseestimate}), we immediately obtain
\begin{eqnarray}
\label{eq:KTT}
-\frac{2}{t}\int_{\Sigma}|K^{TT}|^{2}_{g}\mu_{g}\leq\frac{d}{dt}\int_{\Sigma}|K^{TT}|^{2}_{g}\mu_{g}\leq-\frac{1}{t}\int_{\Sigma}|K^{TT}|^{2}_{g}\mu_{g}.
\end{eqnarray} 
Now integrate (\ref{eq:KTT}) and consider the limit $t\to0$ to yield
\begin{eqnarray}
\frac{\text{const}.}{t}\leq\int_{\Sigma}|K^{TT}|^{2}_{g}\mu_{g}\leq\frac{\text{const}.}{t^{2}}.
\end{eqnarray}
Using the expression of the Dirichlet energy $E[id;\gamma,\rho]$ from equation (\ref{eq:energy}), the following estimate is obtained in the limit $\tau\to-\infty$ i.e., $t\to 0$
\begin{eqnarray}
\label{eq:energy4}
\frac{2 C_{2}}{t}-4\pi\chi \leq E_{\gamma}\leq \frac{2 C_{3}}{t^{2}}-4\pi\chi,
\end{eqnarray}
that implies Dirichlet energy blows up as we approach big-bang singularity, 
$0<C_{2},C_{3}<\infty$. An immediate interpretation of such limiting behavior would be that the corresponding Einstein solution curve leaves every compact set in the Teichm\"uller space (configuration space). This is once again a consequence of the fact that the Dirichlet energy is a proper function on the Teichm\"uller space (see \cite{tromba2012teichmuller} for the detailed proof of the properness of the Dirichlet energy). Therefore, every non-trivial solution curve leaves the Teichm\"uller space at the limit of the big bang. However, we do not know where they converge in the space of projective currents. Note that the space of projective currents is compact and therefore every sequence has a convergent subsequence (since for a metric space, compactness and sequential compactness are equivalent). However, in my context, convergence is a bit more subtle since we are necessarily dealing with curves. We have to extract a sequence $\{l_{\tau_{i}}\}_{i=1}^{\infty}$ and show that it converges in the space of projective currents in the limit $i\to\infty$ ($\lim_{i\to\infty}\tau_{i}=-\infty$) and that the limit does not depend on the choice of the sequence. We want to identify this limit set in the space of projective currents. In fact, we would like to show in the following sections that every non-trivial solution curve indeed attaches to the Thruston boundary of the Teichm\"uller space.}

\section{Asymptotic behavior of the solution curve at big-bang and Thurston boundary}
\noindent In the previous section, we have established that every non-trivial solution curve runs off the edge of the Teichm\"uller space. However, we do not apriori know whether they actually attach to the Thurston boundary. However, when realizing the Teichm\"uller space as a subset of the space of projective currents (which is compact), if we extract a sequence from the solution curve, this must converge somewhere at the limit $\tau\to-\infty$ (after passing to a subsequence and the limit should not depend on the choice of the sequence). Here, we will show that this limit set will be characterized by $\int_{\Sigma_{p}}\sqrt{|k^{TT}|^{2}_{\gamma}}\mu_{\gamma}=C$, $C<\infty$ is a uniform constant. Let us designate this boundary as the `Einstein boundary' of the Teichm\"uller space and denote it by $Ein_{g}$. My goal in this section is to show that this boundary is indeed equivalent to the Thurston boundary that is $\bar{\mathcal{T}\Sigma_{p}}^{Th}=\mathcal{T}\Sigma_{p}\cup \partial\mathcal{T}\Sigma^{Th}_{g}\approx \mathcal{T}\Sigma_{p}\cup Ein_{g}$. Note that Michael Wolf \cite{wolf1989teichmuller} obtained a compactification of Teichm\"uller space through the use of holomorphic quadratic differentials and he proved that his compactification is indeed equivalent to the Thurston compactification. In my case, we are automatically equipped with a holomorphic quadratic differential $k^{TT}$ (the transverse-traceless tensor). However, importantly, Wolf's analysis is quite different from mine (and complementary in nature) in the sense that the Einsteinian dynamics occur in the domain of the associated harmonic map while Wolf's dynamics materialize in the target space. Now we will show the boundedness of $|k^{TT}|^{2}_{\gamma}$ in the limit $\tau\to-\infty$.
\begin{lemma}
\label{theorem2proof1}
Let $(k^{TT},\gamma)$ solve the reduced Einstein equations after imposing the constraints and gauge conditions. Then the following estimates hold for $|k^{TT}|^{2}_{\gamma}$ at the limit $\tau\to-\infty$ (or equivalently $t\to0$)
 \begin{eqnarray}
 \label{eq:kappa}
0<\lim_{t\to0}\int_{\Sigma}\sqrt{|k^{TT}|^{2}_{\gamma}}\mu_{\gamma}\leq C<\infty.
 \end{eqnarray}
\end{lemma}
 \textbf{Proof:} Note that the following entity is conformally invariant 
\begin{eqnarray}
P=\int_{\Sigma_{p}}\sqrt{|K^{TT}|^{2}_{g}}\mu_{g}=\int_{\Sigma_{p}}\sqrt{|k^{TT}|^{2}_{\gamma}}\mu_{\gamma}.
\end{eqnarray} 
Applying the Cauchy-Swartz inequality, Hamiltonian constraint $|K^{TT}|^{2}_{g}=\frac{\tau^{2}}{2}+R(g)$, and time defined in (\ref{eq:time}), we immediately obtain
\begin{eqnarray}
\left(\int_{\Sigma_{p}}\sqrt{|k^{TT}|^{2}_{\gamma}}\nonumber\mu_{\gamma}\right)^{2}=\left(\int_{\Sigma_{p}}\sqrt{|K^{TT}|^{2}_{g}}\mu_{g}\right)^{2}\\\nonumber 
\left(\int_{\Sigma_{p}}\sqrt{|k^{TT}|^{2}_{\gamma}}\nonumber\mu_{\gamma}\right)^{2}\leq\left(\int_{\Sigma_{p}}|K^{TT}|^{2}_{g}\mu_{g}\right)\left(\int_{\Sigma_{p}}\mu_{g}\right)
=\left(\int_{\Sigma_{p}}(\frac{\tau^{2}}{2}+R(g))\mu_{g}\right)\left(\int_{\Sigma_{p}}\mu_{g}\right)\\\nonumber 
=\frac{\tau^{2}}{2}\left(\int_{\Sigma_{p}}\mu_{g}\right)^{2}+4\pi\chi\int_{\Sigma_{p}}\mu_{g}\leq\frac{1}{2t^{2}}V(g)^{2},
\end{eqnarray}
where we have the used Gauss-Bonet theorem $\int_{\Sigma_{p}}R(g)\mu_{g}=4\pi\chi$, where $\chi=2(1-g)<0$ is the Euler characteristic. On the other hand, we know that the volume $V(g)$ of $(\Sigma_{p},g)$ approaches zero at the big bang. However, we will study the evolution of $V(g)$ and obtain a more precise estimate in terms of $|\tau|$. Time differentiating $V(g)=\int_{\Sigma_{p}}\mu_{g}$ (here $p$ in $\Sigma_{p}$ denotes genus while $g$ in $\mu_{g}$ denotes the volume form associated to metric $g$) yields 
\begin{eqnarray}
\frac{dV(g)}{dt}=\frac{1}{2}\int_{\Sigma_{p}}g^{ij}\partial_{t}g_{ij}\mu_{g},
\end{eqnarray}
which together with the evolution equation $\partial_{t}g_{ij}=-2N(K^{TT}_{ij}+\frac{\tau}{2}g_{ij})+(L_{X}g)_{ij}$ yields 
\begin{eqnarray}
\frac{dV(g)}{dt}=\int_{\Sigma_{p}}(-N\tau+\nabla[g]_{i}X^{i})\mu_{g}=-\tau\int_{\Sigma_{p}}N\mu_{g},
\end{eqnarray}
where the total covariant divergence term is dropped following Stokes' theorem. 
Utilizing the estimate of the lapse function $1\leq N\leq2$ (corollary \ref{lapseestimate}) and $t=-\frac{1}{\tau}$ (\ref{eq:time}), we immediately achieve the following bound for the time derivative of the volume $V(g)$
\begin{eqnarray}
\frac{1}{t}V(g)\leq\frac{dV(g)}{dt}\leq\frac{2}{t}V(g),
\end{eqnarray}
integration of which yields the following at the limit $\tau\to-\infty$ or $t\to0$  
\begin{eqnarray}
\text{constant}_{1}\cdot~t^{2} \leq V(g(t))\leq~\text{constant}_{2} \cdot~t
\end{eqnarray}
 Therefore, by using the inequality $0< \left(\int_{\Sigma_{p}}\sqrt{|k^{TT}|^{2}_{\gamma}}\mu_{g}\right)^{2}\leq \frac{1}{2t^{2}}(V(g))^{2}$, we obtain 
\begin{eqnarray}
\label{eq:kappa}
0<\lim_{t\to0}\int_{\Sigma}\sqrt{|k^{TT}|^{2}_{\gamma}}\mu_{\gamma}\leq C<\infty,
\end{eqnarray}
for a uniform constant $C$ (uniform over the conformal structure). Since, $k^{TT}\equiv 0$ implies convergence to a point lying in the interior of $\mathcal{T}\Sigma_{p}$, the left inequality in (\ref{eq:kappa}) is strict (by the blow-up of Dirichlet energy in theorem \ref{theorem1}). This concludes the proof of the lemma.~~~~$\Box$ 

More importantly, $\sup_{\Sigma_{p}}\sqrt{|k^{TT}|^{2}_{\gamma}}(\tau)$ appears explicitly in a later part where we analyze the Gauss-map equation. In that particular analysis, we require a point-wise control of $\sqrt{|k^{TT}|^{2}_{\gamma}}$. Notice that this is the $L^{\infty}$ norm of the holomorphic quadratic differential $\phi:=(k^{TT}_{11}-i k^{TT}_{12})dz^{2}$ with respect to the metric $\gamma$. The obvious problem is that $\phi$ may have singularities on measure zero sets. Remarkably, an integrable holomorphic quadratic differential enjoys the property of possessing at most simple poles at punctures of $\Sigma_{p}$. Now, even though $\Sigma_{p}$ in question does not have punctures, it forms $\delta-$thin regions in the limit of $\tau\to-\infty$. This is precisely the consequence of the blow-up of the Dirichlet energy. If we parameterize the Teichm\"uller space in the space of projective currents by the lengths of $9g-9$ nontrivial elements of $\pi_{1}(\Sigma_{p})$ ($9g-9$ theorem), then the properness of the Dirichlet energy yields geodesics with large hyperbolic length. As a consequence of the Collar lemma \cite{tromba2012teichmuller}, a geodesic transverse to such long geodesic shrinks (again relative to hyperbolic length) leading to the development of a $\delta-$thin region. We will shortly show via the thick-thin decomposition of $\Sigma_{p}$ that an integrable holomorphic quadratic differential is bounded (in the sense of $L^{\infty}$ norm with respect to the metric $\gamma$) even if $\Sigma_{p}$ develops ``bad" parts. Let us first define the norms we are interested in. The $L^{1}$ norm and the $L^{\infty}$ norm (with respect to $\gamma:=e^{2\eta}(dx\otimes dx+dy\otimes dy)$) or Ber's supremum norm of $\phi$ are defined as follows 
\begin{eqnarray}
||\phi||_{L^{1}(\Sigma_{p})}:=\frac{1}{\sqrt{2}}\int_{\Sigma_{p}}|\phi|=\int_{\Sigma_{p}}\sqrt{|k^{TT}|^{2}_{\gamma}}\mu_{\gamma},\\
||\phi||_{L^{\infty}(\Sigma_{p})}:=\sup_{\Sigma_{p}}\sqrt{|k^{TT}|^{2}_{\gamma}}
=\sup_{\Sigma_{p}}\sqrt{\gamma^{ik}\gamma^{jl}k^{TT}_{ij}k^{TT}_{kl}}
=\sup_{\Sigma_{p}}\sqrt{e^{-4\eta}\delta^{ik}\delta^{jl}k^{TT}_{ij}k^{TT}_{kl}}\\\nonumber 
=\sqrt{2}\sup_{\Sigma_{p}}e^{-2\eta}\sqrt{(k^{TT}_{11})^{2}+(k^{TT}_{12})^{2}}.
\end{eqnarray}
Here, we have used the symmetry and traceless property of $k^{TT}$ i.e., $k^{TT}_{12}=k^{TT}_{21}$, and $k^{TT}_{11}+k^{TT}_{22}=0$.
These norms are the natural ones defined for sections of vector bundles defined on $\Sigma_{p}$.
Both norms make the space of holomorphic quadratic differentials on $\Sigma_{p}$ to be a Banach space. Since, the dimension (real) of this space is $6p-6$ (therefore, finite), the $L^{1}$  norm is equivalent to Ber's supremum norm. However, let us explicitly establish the equivalence between $L^{1}$ and $L^{\infty}$ in the case when $\Sigma_{p}$ contains ``bad" parts by invoking the thick-thin decomposition of $\Sigma_{p}$.

Let $\Sigma_{p}$ be a hyperbolic Riemann surface. We will think of $\pi_{1}(\Sigma_{p})$ as the set of the non-trivial loops up to homotopy. Remark that each loop in $\Sigma_{p}$ is
homotopic to a piecewise differentiable loop based at the same point so that we can think of $\pi_{1}(\Sigma_{p})$ as the set of all piecewise differentiable loops up to
homotopy. For $\delta>0$, the thin and thick parts of $\Sigma_{p}$ are defined as follows 
\begin{eqnarray}
 {\Sigma_{p(0,\delta]}:=\{x\in \Sigma_{p}: \exists \langle\alpha\rangle\in\pi_{1}(\Sigma_{p},x)-\{1\}|~l_{\gamma}(\alpha)\leq\delta\}}\\\nonumber 
 {\Sigma_{p[\delta,\infty)}:=\{x\in \Sigma_{p}: \exists \langle\alpha\rangle\in\pi_{1}(\Sigma_{p},x)-\{1\}|~l_{\gamma}(\alpha)\geq\delta\}}.
\end{eqnarray}
Here, $l_{\gamma}(\alpha)$ indicates the length of the geodesic in the homotopy class $\langle\alpha\rangle$ with respect to the hyperbolic metric $\gamma$. The thin part may consist of cusps and Margulis tubes. Since we are dealing with the compact case, the thin part contains a Margulis tube ($\mathbb{S}^{1}\times I$, $I\subset \mathbb{R}$) only. The obvious problem arises in the $\delta-$thin region since, in this region, the length of a geodesic decreases without bound as we approach the big bang. Here we fix a $\delta>0$ and focus on the behavior of the $L^{\infty}$ norm (w.r.t $\gamma_{\tau}$) of the holomorphic quadratic differential ($\phi$) in the $\delta-$thin region since, in the $\delta-$thick region, the $L^{\infty}$ norm is always controlled by the $L^{1}$ norm. We now state two lemmas that conclude the business of controlling the $L^{\infty}$ norm (with respect to $\gamma$) in terms of the $L^{1}$ norm of $\phi$. Note that an integrable holomorphic quadratic differential on a closed (no punctures, no boundary components) Riemann surface does not have poles and therefore has zero principle part. For such an integrable holomorphic quadratic differential on $\Sigma_{p}$ (since it is compact without boundary), the following lemma holds. Proof of this lemma uses results from elementary complex analysis such as the maximum principle for holomorphic functions.\\\\
\begin{proposition}
\label{prop1} \cite{rupflin2013asymptotics}\textit{
For $\delta>0$ and any closed Riemann surface $\Sigma_{p}$, there exists a constant $C<\infty$ depending only on the genus $p$ of $\Sigma_{p}$ and independent of $\delta$ such that for every hyperbolic metric $\gamma$ on $\Sigma_{p}$ the following holds for the holomorphic quadratic differential in the $\delta-$thin region 
\begin{eqnarray}
||\phi||_{L^{\infty}(M_{(0,\delta]})}\leq Ce^{-\pi/\delta}/\delta^{2}||\phi||_{L^{1}(\Sigma_{p})}.
\end{eqnarray}
}
\end{proposition}
Of course, the boundedness follows from the boundedness of $e^{-\pi/\delta}/\delta^{2}$. In the $\delta-$thick region, $L^{\infty}$ (with respect to the metric $\gamma$) control in terms of $L^{1}$ (with respect to the metric $\gamma$) is trivial and follows from the following lemma.\\
\begin{proposition}
\label{prop2}\cite{rupflin2013asymptotics}
 \textit{For any $\delta>0$ and any closed Riemann surface $\Sigma_{p}$, there exists a constant $C_{\delta}<\infty$ depending only on $\delta$ and the genus $g$ of $\Sigma_{p}$ such that for every hyperbolic metric $\gamma$ on $\Sigma_{p}$ the following holds for the holomorphic quadratic differential in the $\delta-$thick region 
\begin{eqnarray}
||\phi||_{L^{\infty}(M_{[\delta,\infty)})}\leq C_{\delta}||\phi||_{L^{1}(\Sigma_{p})}.
\end{eqnarray}}
\end{proposition}
Now consider the case when $\Sigma_{p}$ completely degenerates and forms punctures (From the Deligne-Mumford compactness theorem, punctured surfaces are achieved as a limit of a Riemann surface degenerating via collapsing non-trivial simply closed geodesics). Now, the analysis becomes a little more subtle since integrable holomorphic quadratic differentials may have a simple pole at a puncture (at worst). A neighborhood of this puncture corresponds to a cusp and is equivalent to a punctured open disc (punctured at $0$) equipped with the metric $e^{2\eta}(dx^{2}+dy^{2})=\frac{1}{|z|^{2}(\log(|z|))^{2}}|dz|^{2}$. Now, roughly it is clear that the simple pole of $\phi$ cancels in the norm $||\phi||_{L^{\infty}}:=e^{-2\eta}\sqrt{k^{2}_{11}+k^{2}_{12}}$. But for completeness, we state the following proposition from \cite{rupflin2013asymptotics}.\\
\begin{proposition}
\label{prop3}
Let $(\Sigma_{p},\gamma)$ be a hyperbolic Riemann surface with finite area. Then for any holomorphic quadratic differential $\phi$ of $\Sigma_{p}$ the following are equivalent\\
1. $||\phi||_{L^{1}(\Sigma_{p})}\leq C, C<\infty\\
2. ||\phi||_{L^{\infty}(\Sigma_{p})}\leq C, C<\infty$\\
3. At each of the punctures of $\Sigma_{p}$ the differential $\phi$ has at worst a simple pole.
\end{proposition}

\noindent Now let us consider a sequence $\{(\gamma_{\tau_{j}},\phi_{\tau_{j}},\tau_{j}\}_{i=1}^{\infty}$ lying on the solution curve of the Einstein flow (on the phase space) with $\lim_{j\to\infty}\tau_{j}=-\infty$. If each member of the sequence satisfies $||\phi_{\tau_{j}}||_{L^{1}(\Sigma_{p})}\leq C$ with the limit satisfying $\lim_{j\to\infty}||\phi_{\tau_{j}}||_{L^{1}(\Sigma_{p})}\leq C$ (for a uniform $C$; this is precisely what we have proved in lemma (\ref{theorem2proof1}), then from the propositions (\ref{prop1}-\ref{prop3}) we conclude that the $L^{\infty}$ norm of the limit is also bounded, that is, the following is satisfied 
\begin{eqnarray}
\label{eq:estimate4}
\lim_{j\to\infty}\sup_{\Sigma_{p}}\sqrt{|k^{TT}|^{2}_{\gamma_{\tau_{j}}}}\leq C, C<\infty,
\end{eqnarray}
(note that from the uniqueness property of the Cauchy problem associated with the vacuum Einstein's equations indicate that if the limit $\lim_{j\to\infty}\sup_{\Sigma_{p}}\sqrt{|k^{TT}|^{2}_{\gamma_{\tau_{j}}}}$ exists, it must be unique).
Now we go back to the following point-wise inequality
(\ref{eq:estimate1})
\begin{eqnarray}
\label{eq:estimate8}
\frac{2}{\tau^{2}}\leq e^{2\varphi}\leq \frac{1+\sqrt{1+2\tau^{2}\sup_{\Sigma_{p}}|k^{TT}|^{2}_{\gamma}(\tau)}}{\tau^{2}}.
\end{eqnarray}
Using the fact that $\sup_{\Sigma_{p}}\sqrt{|k^{TT}|^{2}_{\gamma}}\leq C_{\infty}<\infty$, we may conclude that the following estimate of $\sup_{\Sigma_{p}}e^{2\varphi}$ holds in the limit $\tau\to-\infty$  
\begin{eqnarray}
\frac{2}{\tau^{2}}\leq e^{2\varphi}\leq \frac{C_{\varphi}}{|\tau|},
\end{eqnarray}
for a suitable constant $0<C_{\varphi}<\infty$. We shall prove an improved version of these estimates later in lemma \ref{theoremproof2}.
We have now obtained the necessary estimates from Einstein's equations in the CMCSH gauge. Utilizing these estimates we want to establish a relation between the hyperbolic length of a nontrivial element of $\pi_{1}(\Sigma_{p})$ and its transverse measure against the measured foliation associated with the holomorphic quadratic differential. As mentioned previously, we have a natural holomorphic quadratic differential associated to the Einstein flow due to the fact that corresponding to each transverse-traceless tensor $k^{TT}$, we may associate a holomorphic quadratic differential. Here, we define the following quadratic differential
 \begin{eqnarray}
\label{eq:defnquad}
\phi=\phi(z)dz^{2}=(k^{TT}_{11}-\sqrt{-1} k^{TT}_{12})dz^{2}
 =k^{TT}_{11}(dx^{2}-dy^{2})+2k^{TT}_{12}dxdy\\\nonumber
 -\sqrt{-1}(k^{TT}_{12}(dx^{2}-dy^{2})-2k^{TT}_{11}dxdy)
 =k+\sqrt{-1}\xi.
 \end{eqnarray}
Note that the transverse-traceless tensor $k^{TT}$ may be recovered as follows
 \begin{eqnarray}
k^{TT}=\mathcal{R}\left(\phi(z)dz^{2}\right).
 \end{eqnarray}
 The transverse-traceless property of $k^{TT}_{ij}$ precisely implies $\frac{\partial \phi}{\partial \bar{z}}=0$ i.e., $\phi$ is holomorphic. This establishes the well-known homeomorphism between the space of holomorphic quadratic differentials and the space of transverse-traceless tensors. In addition, we have a natural homeomorphism between the space of transverse traceless tensors on $(\Sigma_{p},\gamma)$ and the Teichm\"uller space from the Einstein flow (for a detailed analysis see \cite{moncrief2007relativistic}).  Once we have a quadratic differential we immediately obtain horizontal and vertical measured foliations associated with this holomorphic quadratic differential. The transverse measures of a non-trivial element of $\pi_{1}(\Sigma_{p})$ with respect to the vertical measured foliation and horizontal measured foliation are (as follows from (\ref{eq:VM}) and ($\ref{eq:HM}$)) 
 \begin{eqnarray}
 \mu_{vert}(\mathcal{C})&=&\oint_{\mathcal{C}}\sqrt{\frac{k+\sqrt{k^{2}+\xi^{2}}}{2}},\\
  \mu_{hor}(\mathcal{C})&=&\oint_{\mathcal{C}}\sqrt{\frac{\sqrt{k^{2}+\xi^{2}}-k}{2}},
 \end{eqnarray}
 respectively.
Let us consider that the tangent vector field to the curve $\mathcal{C}$ be $u^{1}\frac{\partial}{\partial x^{1}}+u^{2}\frac{\partial}{\partial x^{2}}$ and denote this by $(u^{1},u^{2})^{T}$. The term $k^{2}$ may be written as the bi-linear form $k^{TT}_{ij}u^{i}u^{j}(d\lambda)^{2}$, where $\lambda$ is the parameter along $\mathcal{C}$. Similarly, the term $\xi$ may be written as $k^{TT}_{im}J^{m}_{j}u^{i}u^{j}=k^{TT}_{im}u^{i}v^{m}$, where $J=\left[\begin{array}{cc}
0 &1\\
-1 &0
\end{array}
\right],$ and $v^{m}=J^{m}_{j}u^{j}$ that is $v=(-u^{2},u^{1})^{T}$. More importantly, we see that the following holds in isothermal coordinates ($\gamma=\gamma(z)|dz|^{2}, \gamma(z)=e^{\delta(z)}, \delta(z):\Sigma_{p}\to \mathbb{R}$)
\begin{eqnarray}
\label{eq:orth}
\gamma(u,v)=\gamma(z)(-u^{1}u^{2}+u^{2}u^{1})=0,
\end{eqnarray}
that is, $u$ and $v$ are orthogonal to each other with respect to the metric $\gamma$. This is precisely a consequence of the existence of an isothermal chart around any point on $\Sigma_{p}$ and since $\gamma(u,v)$ is a scalar, vanishing in one coordinate chart implies vanishing in every coordinate chart (as mentioned in the beginning, we use isothermal coordinates throughout). The transverse measure to the vertical foliation may be written as follows
\begin{eqnarray}
\label{eq:measure}
\mu_{vert}(\mathcal{C})&=&\oint_{\mathcal{C}}\sqrt{\Bigl|\frac{k^{TT}_{ij}u^{i}u^{j}+\sqrt{(k^{TT}_{ij}u^{i}u^{j})^{2}+(k^{TT}_{ij}u^{i}v^{j})^{2}}}{2}\Bigr|}d\lambda.
\end{eqnarray} 
Notice that these measures are diffeomorphism (of $\Sigma_{p})$ and re-parametrization (of the curve $\mathcal{C}$) invariant. 

Let us understand the limiting behavior of the transverse measure of a closed homotopically non-trivial curve $\mathcal{C}$ with respect to the vertical foliation at a heuristic level. We now compute the $\gamma-$length of a geodesic in the homotopy class $[\mathcal{C}]$ and relate it to its transverse measure associated with the measured foliation of the holomorphic quadratic differential $\phi$. Through the unique solution of the Monge-Ampere equation, the Gauss map equation defines a ray structure of the Einstein equations. Therefore analyzing the asymptotic behavior of the Monge-Ampere equation is in principle the same as analyzing the Gauss map equation while satisfying Einstein's equations through the associated Hamilton-Jacobi equation. In addition, analysis of the Gauss map equation seems more tractable (and relevant) because we have a handful of estimates from the elliptic equations associated with the Einstein dynamics. Using the Gauss map equation, we obtain 
\begin{eqnarray}
\rho_{ij}u^{i}u^{j}=|K|^{2}_{g}g_{ij}u^{i}u^{j}+2\tau K_{ij}u^{i}u^{j}-\tau^{2}g_{ij}u^{i}u^{j},\\\nonumber
=(|K^{TT}|^{2}_{g}+\frac{\tau^{2}}{2})g_{ij}u^{i}u^{j}+2\tau K^{TT}_{ij}u^{i}u^{j}\\\nonumber
=(e^{-4\varphi}|k^{TT}|^{2}_{\gamma}+\frac{\tau^{2}}{2})e^{2\varphi}\gamma_{ij}u^{i}u^{j}
+2\tau k^{TT}_{ij}u^{i}u^{j}.
\end{eqnarray}

\noindent We do know the fact that $\rho\in\mathcal{T}\Sigma_{p}$ is fixed and the $\rho-$length of $\mathcal{C}$ is bounded (due to the properness of the Dirichlet energy which remains finite in the interior of the Teichm\"uller space). Following the Gauss map equation, we have the following 
\begin{eqnarray}
|\rho_{ij}u^{i}u^{j}|&=&|\left\{|K^{TT}|^{2}_{g}+\frac{\tau^{2}}{2}\right\}g_{ij}u^{i}u^{j}+2\tau k^{TT}_{ij}u^{i}u^{j}|,\\\nonumber
&\geq&|\left\{|K^{TT}|^{2}_{g}+\frac{\tau^{2}}{2}\right\}g_{ij}u^{i}u^{j}|-2|\tau k^{TT}_{ij}u^{i}u^{j}|,\\\nonumber
&=&||k^{TT}|^{2}_{\gamma}e^{-2\varphi}\gamma_{ij}u^{i}u^{j}+\frac{\tau^{2}}{2}e^{2\varphi}\gamma_{ij}u^{i}u^{j}|-2|\tau k^{TT}_{ij}u^{i}u^{j}|,\\\nonumber
&\geq&2|\sqrt{\frac{|k^{TT}|^{2}_{\gamma}\tau^{2}}{2}(\gamma_{ij}u^{i}u^{j})^{2}}|-2|\tau k^{TT}_{ij}u^{i}u^{j}|,\nonumber
\end{eqnarray}
that is, 
\begin{eqnarray}
\frac{1}{\sqrt{2}}\sqrt{|k^{TT}|^{2}_{\gamma}}\gamma_{ij}u^{i}u^{j}\leq |k^{TT}_{ij}u^{i}u^{j}|+\frac{1}{2|\tau|}\rho_{ij}u^{i}u^{j}.
\end{eqnarray}
Here, we have used $l^{2}+m^{2}\geq 2lm$ for $l,m\in\mathbb{R}$. Now point-wise norm of $\sqrt{|k^{TT}|^{2}_{\gamma}}$ satisfies
\begin{eqnarray}
0\leq |k^{TT}|^{2}_{\gamma} \leq C^{2}_{\infty}. 
\end{eqnarray}
Notice that the infimum of $|k^{TT}|^{2}_{\gamma}$ may be zero since the holomorphic quadratic differential $\phi:=(k^{TT}_{11}-ik^{TT}_{12})dz^{2}$ has finite number of zeros. Let us consider that the infimum of $\sqrt{|k^{TT}|^{2}_{\gamma}}$ be $C_{f}$ which is strictly positive provided that we stay away from the zeros (a finite number) of the quadratic differential $\phi$ (which correspond to the singularities of the associated measured foliation). Let us consider that the quadratic differential has zeros at $(z_{1}, z_{2},....,z_{n}),$ $n<\infty$. Consider $\epsilon$ disks $D_{\epsilon}(z_{i})$ around each of the zeros. As these zeros correspond to the singularities of the associated measured foliation, we will consider the transverse measure on $\Sigma^{'}_{p}=\Sigma_{p}-\{\cup_{i=1}^{n}D_{\epsilon}(z_{i})\}$ (a detailed rationale is sketched in Wolf's article and therefore we do not repeat the same here). On $\Sigma^{'}_{p}$, the previous inequality becomes  
\begin{eqnarray}
\frac{|C_{f}|}{\sqrt{2}}\gamma_{ij}u^{i}u^{j}\leq |k^{TT}_{ij}u^{i}u^{j}|+\frac{1}{2|\tau|}\rho_{ij}u^{i}u^{j},\\
\lim_{\tau\to-\infty}\frac{|C_{f}|}{\sqrt{2}}\gamma_{ij}u^{i}u^{j}\leq \lim_{\tau\to-\infty}|k^{TT}_{ij}u^{i}u^{j}|+\lim_{\tau\to-\infty}\frac{1}{2|\tau|}\rho_{ij}u^{i}u^{j}.
\end{eqnarray}
Now notice the fact that $\rho-$length of $\mathcal{C}$ is finite and independent of $\tau$ since $\rho$ lies in the interior of the Teichm\"uller space and therefore 
\begin{eqnarray}
\lim_{\tau\to-\infty}\frac{1}{2|\tau|}\rho_{ij}u^{i}u^{j}=0.
\end{eqnarray}
We obtain the following inequality
\begin{eqnarray}
\lim_{\tau\to-\infty}\frac{|C_{f}|}{\sqrt{2}}\gamma_{ij}u^{i}u^{j}\leq \lim_{\tau\to-\infty}|k^{TT}_{ij}u^{i}u^{j}|.
\end{eqnarray}
Let us analyze the Gauss-map equation in a different way  
\begin{eqnarray}
|\rho_{ij}u^{i}u^{j}-2\tau k^{TT}_{ij}u^{i}u^{j}|&=&\Bigl| |(K^{TT}|^{2}_{g}+\frac{\tau^{2}}{2})g_{ij}u^{i}u^{j}\Bigr|.
\end{eqnarray}
Now utilizng the estimate of $|K^{TT}|^{2}_{g}$ from (\ref{eq:estimate1}), we obtain 
\begin{eqnarray}
|2\tau k^{TT}_{ij}u^{i}u^{j}|-|\rho_{ij}u^{i}u^{j}|
&\leq&\tau^{2}e^{2\varphi}\gamma_{ij}u^{i}u^{j},
\end{eqnarray}
which utilizing the estimate (\ref{eq:estimate1}) yields
\begin{eqnarray}
|2\tau k^{TT}_{ij}u^{i}u^{j}|-|\rho_{ij}u^{i}u^{j}|&\leq&\left(1+\sqrt{1+2\tau^{2}\sup_{x\in\Sigma^{'}_{g}}|k^{TT}|^{2}_{\gamma}(\tau)}\right)\gamma_{ij}u^{i}u^{j}.
\end{eqnarray}
Substituting the estimate (\ref{eq:estimate4}) into the previous inequality leads to 
\begin{eqnarray}
|k^{TT}_{ij}u^{i}u^{j}|\leq \left(\frac{1}{2\tau}+\frac{\sqrt{1+2C_{\infty}^{2}\tau^{2}}}{2\tau}\right)\gamma_{ij}u^{i}u^{j}+\frac{1}{2|\tau|}\rho_{ij}u^{i}u^{j},\nonumber
\end{eqnarray}
and therefore, in the limit $\tau\to-\infty$
\begin{eqnarray}
\lim_{\tau\to-\infty}\frac{|k^{TT}_{ij}u^{i}u^{j}|}{\gamma_{ij}u^{i}u^{j}}&\leq&\lim_{\tau\to-\infty}\left(\frac{1}{2\tau}+\frac{\sqrt{1+2C_{\infty}^{2}\tau^{2}}}{2\tau}\right)\\\nonumber
&=&\frac{|C_{\infty}|}{\sqrt{2}}.
\end{eqnarray}
In a sense, we have as $\tau\to-\infty$
\begin{eqnarray}
\frac{|C_{f}|}{\sqrt{2}}\gamma_{ij}u^{i}u^{j}\leq|k^{TT}_{ij}u^{i}u^{j}|\leq \frac{|C_{\infty}|}{\sqrt{2}}\gamma_{ij}u^{i}u^{j},
\end{eqnarray}
with $0<C^{2}_{f}<C_{\infty}^{2}<\infty$.
This is an important expression obtained at the limit of the big-bang ($\tau\to-\infty$). On the other hand, the expression for the transverse measure of the vertical foliation reads (\ref{eq:measure})
\begin{eqnarray}
\mu_{vert}(\mathcal{C})&=&\oint_{\mathcal{C}}\sqrt{\Bigl|\frac{k^{TT}_{ij}u^{i}u^{j}+\sqrt{(k^{TT}_{ij}u^{i}u^{j})^{2}+(k^{TT}_{ij}u^{i}v^{j})^{2}}}{2}\Bigr|}d\lambda.
\end{eqnarray}
We still need to obtain an estimate for the term $k^{TT}u^{i}v^{j}$. In a sense, the Einstein flow drives the solution curve in such a way that the measured foliation behaves in this way at the limit of the big-bang singularity. 
Therefore, we obtain the following crucial relation in the big-bang limit ($\tau\to-\infty$)
\begin{eqnarray}
\label{eq:proportional}
\mu_{vert}(\mathcal{C})= C\oint_{\mathcal{C}}\sqrt{\gamma_{ij}u^{i}u^{j}}d\lambda=Cl_{\gamma}(\mathcal{C}),
\end{eqnarray}
for a suitable constant $\frac{|C_{f}|^{1/2}}{2^{1/4}}\leq C\leq \frac{|C_{\infty}|^{1/2}}{2^{1/4}}$. An important point to notice is that the constants $C_{f}$ and $C_{\infty}$ are uniform in the sense that they do not depend on the chosen homotopy class $[\mathcal{C}]$. Now this does not imply that $C$ is independent of the homotopy class chosen. In fact, we need to show that $C$ does not depend on the homotopy class of loops at the limit $\tau\to-\infty$. This follows since we will show that $|k^{TT}|^{2}_{\gamma}$ behaves as a constant modulo factor involving inverse power of the mean extrinsic curvature $\tau$ as $\tau$ approaches $-\infty$ (i.e., big bang). We claim the following.\\
\begin{lemma}
\label{theoremproof2}
The following solve Einstein's evolution and constraint equations as $\tau\to-\infty$
\begin{eqnarray}
|k^{TT}|^{2}_{\gamma}=a^{2}-O(\frac{1}{|\tau|^{2}})~a.e~on~\Sigma_{p}~as~\tau\to-\infty,\\
\label{eq:confsolution}
e^{2\varphi}=\frac{\sqrt{2}a}{|\tau|}+\frac{1}{\tau^{2}}+O(\frac{1}{|\tau|^{3}}),\\
N=1+\frac{1}{\sqrt{2}|\tau|a}+O(\frac{1}{\tau^{2}})~on~\Sigma_{p}~as~\tau\to-\infty,\\
\gamma(X,X)=O(\frac{1}{|\tau|})~a.e~on~ \Sigma_{p}~as~\tau\to-\infty.
\end{eqnarray}
where $a^{2}>0$ is a universal constant that may only depend on $\chi(\Sigma_{p})$.
\end{lemma}
\begin{proof}
Firstly, $a\neq 0$ since that would imply $k^{TT}\to 0$ as $\tau\to-\infty$. But this would imply that the corresponding metric $\gamma$ lies in the interior of the Teichm\"uller space which is a contradiction to the fact that Dirichlet energy blows up as $\tau\to-\infty$ (theorem 1.1). Explicit calculation using $|k^{TT}|^{2}_{g}=g^{ij}g^{kl}k^{TT}_{ik}k^{TT}_{jl}$ and the Einstein evolution equations (\ref{eq:evol1}-\ref{eq:evol2}) yields the desired evolution equation for $|k^{TT}|^{2}_{g}$   
\begin{eqnarray}
\partial_{t}|k^{TT}|^{2}_{g}=L_{X}|k^{TT}|^{2}_{g}+2N\tau |k^{TT}|^{2}_{g}+2\nabla[g]^{i}\nabla[g]^{j}Nk^{TT}_{ij}.
\end{eqnarray}
Now utilizing the conformal transformation $g_{ij}=e^{2\varphi}\gamma_{ij}$ and noting $k^{TT}_{ij}$ is conformally invariant, the evolution equation for $|k^{TT}|^{2}_{g}$ may be transformed into an evolution equation for $|k^{TT}|^{2}_{\gamma}$ 
\begin{eqnarray}
\label{eq:evolnew}
\partial_{t}|k^{TT}|^{2}_{\gamma}+e^{4\varphi}|k^{TT}|^{2}_{\gamma}\partial_{t}e^{-4\varphi}=2N\tau |k^{TT}|^{2}_{\gamma}+L_{X}|k^{TT}|^{2}_{\gamma}-e^{4\varphi}L_{X}e^{-4\varphi}\\\nonumber 
+\gamma^{ik}\gamma^{jl}\nabla[\gamma]_{i}\nabla_{j}Nk^{TT}_{kl}-\frac{1}{2}e^{-2\varphi}(\gamma^{ik}\gamma^{ml}\partial_{i}e^{2\varphi}+\gamma^{mk}\gamma^{jl}\partial_{j}e^{2\varphi})\nabla_{m}Nk^{TT}_{kl},
\end{eqnarray}
where we have utilized the identity
\begin{eqnarray}
\nabla[g]_{i}\nabla_{j}N=\nabla[\gamma]_{i}\nabla_{j}N-\frac{1}{2}g^{mn}(\nabla[\gamma]_{i}g_{nj}+\nabla[\gamma]_{j}g_{in}-\nabla[\gamma]_{n}g_{il}).
\end{eqnarray}
Using  
\begin{eqnarray}
|k^{TT}|^{2}_{\gamma}=a^{2}-O(\frac{1}{\tau^{2}}),\\   
N=1+\frac{1}{\sqrt{2}|\tau|a}+O(\frac{1}{\tau^{2}}),\\
e^{2\varphi}=\frac{\sqrt{2}a}{|\tau|}+\frac{1}{\tau^{2}}+O(\frac{1}{|\tau|^{3}}),\\
e^{-2\varphi}=\frac{|\tau|}{\sqrt{2}a}-\frac{1}{2a^{2}}+O(\frac{1}{|\tau|})\\
\gamma(X,X)=O(\frac{1}{|\tau|}),
\end{eqnarray}
We may compute each term of the evolution equation (\ref{eq:evolnew}) as follows 
\begin{eqnarray}
e^{4\varphi}|k^{TT}|^{2}_{\gamma}\partial_{t}e^{-4\varphi}
=\left((\frac{2a^{2}}{\tau^{2}}+\frac{1}{\tau^{4}}+\frac{2\sqrt{2}a}{\tau^{3}})+O(\frac{1}{\tau^{4}})\right)(a^{2}+O(\frac{1}{\tau^{2}}))(\frac{\tau^{3}}{a^{2}}\nonumber-\frac{\tau^{2}}{\sqrt{2}a^{3}}+O(|\tau|))\\\nonumber 
=2a^{2}\tau+\sqrt{2}a+O(\frac{1}{|\tau|}),\\\nonumber
2N\tau|k^{TT}|^{2}_{\gamma}=2\tau a^{2}+\sqrt{2}a+O(\frac{1}{|\tau|}),~
e^{4\varphi}L_{X}e^{-4\varphi}=O(\frac{1}{\tau^{3}}),\\\nonumber
\gamma^{ik}\gamma^{jl}\nabla[\gamma]_{i}\nabla_{j}Nk^{TT}_{kl}=O(\frac{1}{\tau^{2}}),\\\nonumber
\frac{1}{2}e^{-2\varphi}(\gamma^{ik}\gamma^{ml}\partial_{i}e^{2\varphi}+\gamma^{mk}\gamma^{jl}\partial_{j}e^{2\varphi})\nabla_{m}Nk^{TT}_{kl}=O(\frac{1}{\tau^{4}}).
\end{eqnarray}
After substituting into the evolution equation (\ref{eq:evolnew}), we observe that the dangerous terms ($O(|\tau|)$ and $O(1)$ terms) are precisely canceled with their respective negative counterparts. Therefore we observe that the evolution equation for $|k^{TT}|^{2}_{\gamma}$ is solved by $|k^{TT}|^{2}_{\gamma}=a^{2}-O(\frac{1}{\tau^{2}})$ almost everywhere on $\Sigma_{p}$ as $\tau\to-\infty$.
Substitute this form of $e^{2\varphi}$ in the Lichnerowicz equation (\ref{eq:constraint}) to yield  
\begin{eqnarray}
-2\Delta_{\gamma}\varphi+e^{-2\varphi}(|k^{TT}|^{2}_{\gamma}-a^{2})=O(\frac{1}{|\tau|})
\end{eqnarray}
integration of which yields at $\tau\to-\infty$ 
\begin{eqnarray}
\int_{\Sigma}e^{-2\varphi}(|k^{TT}|^{2}_{\gamma}-a^{2})\mu_{\gamma}=O(\frac{1}{|\tau|}).
\end{eqnarray}
and therefore 
\begin{eqnarray}
\label{eq:constant}
|k^{TT}|^{2}_{\gamma}=a^{2}-O(\frac{1}{|\tau|^{2}})~a.e~on~\Sigma~as~\tau\to-\infty
\end{eqnarray}
solves the Lichnerowicz equation.  
Substituting $|k^{TT}|^{2}_{\gamma}=a^{2}+O(\frac{1}{|\tau|^{2}})~a.e~on~\Sigma~as~\tau\to-\infty$ into the lapse equation yields
\begin{eqnarray}
N=1+\frac{1}{|\sqrt{2}\tau|a}+O(\frac{1}{\tau^2})~as~\tau\to-\infty
\end{eqnarray}
which yields through the shift equation and integration by parts argument (\ref{eq:shift})
\begin{eqnarray}
\int_{\Sigma_{p}}(\nabla[\gamma]_{i}X^{j}\nabla[\gamma]_{k}X^{l}\gamma^{ik}\gamma_{jl}+\gamma_{ij}X^{i}X^{j})\mu_{\gamma}=O(\frac{1}{\tau^{2}})
\end{eqnarray}
yielding
\begin{eqnarray}
\gamma(X,X)=O(\frac{1}{\tau^{2}})~a.e~on~\Sigma_{p}~as~\tau\to-\infty.
\end{eqnarray}
\end{proof}

\begin{remark}
In Wolf's study \cite{wolf1989teichmuller}, the constant $a$ is $\frac{\sqrt{2}}{4}$. We could explicitly obtain this numerical constant by studying the identity $\int_{\Sigma_{p}}R(g)\mu_{g}=4\pi\chi(\Sigma_{p})$. However, we do not need the exact value of $a$ in my context and therefore omit such calculations. 
\end{remark}

\noindent This property is extremely important and indicates an asymptotically velocity-dominated behavior i.e., the evolution equations are effective ordinary differential equations in time as one approaches the big-bang since the spatial parts are weighted by the inverse power of the mean curvature. Velocity term dominated behavior (VTD) has also been previously noted in the context of big-bang singularity \cite{VDT}. Since $|k^{TT}|^{2}_{\gamma}$ asymptotically approaches a constant $a^{2}$, we obtain at the limit $\tau\to-\infty$, $C_{\infty}=C_{f}=a$ and therefore the constant $C$ in equation (\ref{eq:proportional}) is independent of the homotopy class of curve i.e.,  $C=\frac{a^{\frac{1}{2}}}{2^{1/4}}=$constant on $\Sigma_{p}$ as $\tau\to-\infty$.\\
\begin{lemma}
\label{theorem2proof3}
Let $\{\gamma_{j},k^{TT}_{j},\tau_{j}\}$ be any sequence such that $\gamma_{j}$ leaves every compact set of $\mathcal{T}\Sigma_{p}$,~ $\tau_{j}\to-\infty$ as $j\to\infty$. Then for all $[\mathcal{C}]\in\mathcal{A}$ and for $j$ sufficiently large, there exists $K_{1},K_{2}>0$ both depending on $\gamma_{j}$ and $[\mathcal{C}]$, such that
\begin{eqnarray}
\label{eq:lengthfoliation}
K_{1}l_{\gamma_{j}}(\mathcal{C})\leq \mu_{vert_{j}}(\mathcal{C})\leq K_{2}l_{\gamma_{j}}(\mathcal{C})
\end{eqnarray}
where the constant $K_{1}, K_{2}\to\frac{a^{\frac{1}{2}}}{2^{\frac{1}{4}}}$ as $j\to\infty$. 
\end{lemma}
\textbf{Proof:} Let us define the following entities 
\begin{eqnarray}
\mathcal{I}n(\tau):=\inf_{\Sigma_{p}^{'}}\sqrt{|k^{TT}|^{2}_{\gamma}(\tau)}\\
\mathcal{S}(\tau):=\sup_{\Sigma_{p}^{'}}\sqrt{|k^{TT}|^{2}_{\gamma}(\tau)}.
\end{eqnarray}
Both $\mathcal{I}n(\tau)$ and $\mathcal{S}(\tau)$ are continuous functions of $\tau$ by existence-uniqueness-continuity (or well-posedness) of the Einstein's equations \cite{anderson1997global}. Clearly the following holds by continuity and the result of lemma 7
\begin{eqnarray}
\lim_{j\to\infty}\inf\mathcal{I}n(\tau_{j})=a,~
\lim_{j\to\infty}\sup\mathcal{S}(\tau_{j})=a.
\end{eqnarray}
Gauss-map equation yields
\begin{eqnarray}
\label{eq:tau1}
\frac{1}{\sqrt{2}}\mathcal{I}n(\tau_{j})\gamma_{j}(u,u)\leq |k^{TT}_{j}(u,u)|+\frac{1}{|\tau_{j}|}\rho(u,u)\\
\label{eq:tau2}
|k^{TT}_{j}(u,u)|\leq \left(\frac{1}{2|\tau_{j}|}+\frac{\sqrt{1+2\mathcal{S}(\tau_{j})^{2}\tau^{2}_{j}}}{2|\tau_{j}|}\right)\gamma_{j}(u,u)+\frac{1}{|\tau_{j}|}\rho(u,u)\\
\label{eq:linequal1}
|k^{TT}_{j}(u,v)|\leq \frac{1}{|\tau_{j}|}|\rho(u,v)|,
\end{eqnarray}
where the metric $\rho$ is fixed i.e., independent of $\tau$ and lies in the interior of the Teichm\"uller space and $u\perp v$. Consider $|k^{TT}_{j}(u,u)|>\delta$ as $j\to\infty$ for a fixed $\delta>0$ (If $|k^{TT}_{j}(u,u)|\to 0$ as $j\infty$, then (\ref{eq:tau1}) and (\ref{eq:tau2}) are trivially satisfied). We write 
\begin{eqnarray}
\label{eq:tau3}
\oint_{\mathcal{C}}\sqrt{|k^{TT}_{j}(u,u)|}\leq \sup_{\mathcal{C}}\sqrt{\frac{1+2\rho(u,u)}{2|\tau_{j}|\gamma_{j}(u,u)}+\frac{\sqrt{1+2\mathcal{S}(\tau_{j})^{2}\tau^{2}_{j}}}{2|\tau_{j}|}}\oint_{\mathcal{C}}\sqrt{\gamma_{j}(u,u)}
\end{eqnarray}
and 
\begin{eqnarray}
\label{eq:tau4}
\oint_{\mathcal{C}}\sqrt{|k^{TT}_{j}(u,u)|}\geq \inf_{\mathcal{C}} \sqrt{\frac{\mathcal{I}n(\tau_{j})}{\sqrt{2}}-\frac{\rho(u,u)}{|\tau_{j}|\gamma_{j}(u,u)}}\oint_{\mathcal{C}}\sqrt{\gamma_{j}(u,u)}.
\end{eqnarray}
Now go back to the formula for the transverse measure to the vertical foliation and obtain the following inequality 
\begin{eqnarray}
\oint_{\mathcal{C}}\sqrt{k^{TT}_{j}(u,u)}\leq \mu_{vert_{j}}([\mathcal{C}])=
\oint_{\mathcal{C}}\sqrt{|\frac{|k^{TT}_{j}(u,u)|+\sqrt{[k^{TT}_{j}(u,u)]^{2}+[k^{TT}_{j}(u,v)]^{2}}}{2}|}\nonumber
\end{eqnarray}
since $[k^{TT}_{j}(u,v)]^{2}\geq 0$. Now utilizing (\ref{eq:linequal1}), we obtain
\begin{eqnarray}
\label{eq:inequalityneeded}
\oint_{\mathcal{C}}\sqrt{k^{TT}_{j}(u,u)}\leq \mu_{vert_{j}}([\mathcal{C}])\leq
\oint_{\mathcal{C}}\sqrt{|\frac{|k^{TT}_{j}(u,u)|+\sqrt{[k^{TT}_{j}(u,u)]^{2}+[\frac{\rho(u,v)}{|\tau_{j}|}]^{2}}}{2}|}\\\nonumber
\leq \sup_{\mathcal{C}}\sqrt{|\frac{1+\sqrt{1+\frac{|\rho(u,u)|^{2}}{|\tau_{j}|^{2}|k^{TT}_{j}(u,u)|^{2}}}}{2}|}\oint_{\mathcal{C}}\sqrt{|k^{TT}_{j}(u,u)|}
\end{eqnarray}
 which together with (\ref{eq:tau3}) and (\ref{eq:tau4}) yields 
 \begin{eqnarray}
\inf_{\mathcal{C}} \sqrt{\frac{\mathcal{I}n(\tau_{j})}{\sqrt{2}}-\frac{\rho(u,u)}{|\tau_{j}|\gamma_{j}(u,u)}}\oint_{\mathcal{C}}\sqrt{\gamma_{j}(u,u)}\leq \mu_{vert_{j}}([\mathcal{C}])\\\nonumber 
 \leq \sup_{\mathcal{C}}\sqrt{|\frac{1+\sqrt{1+\frac{|\rho(u,u)|^{2}}{|\tau_{j}|^{2}|k^{TT}_{j}(u,u)|^{2}}}}{2}|}\sup_{\mathcal{C}}\sqrt{\frac{1+2\rho(u,u)}{2|\tau_{j}|\gamma_{j}(u,u)}\nonumber+\frac{\sqrt{1+2\mathcal{S}(\tau_{j})^{2}\tau^{2}_{j}}}{2|\tau_{j}|}}\oint_{\mathcal{C}}\sqrt{\gamma_{j}(u,u)}.
 \end{eqnarray}
Therefore 
\begin{eqnarray}
K_{1}l_{\gamma_{j}}(\mathcal{C})\leq \mu_{vert_{j}}(\mathcal{C})\leq K_{2}l_{\gamma_{j}}(\mathcal{C}),
\end{eqnarray}
where $K_{1}:=\inf_{\mathcal{C}}\sqrt{\frac{\mathcal{I}n(\tau_{j})}{\sqrt{2}}-\frac{\rho(u,u)}{|\tau_{j}|\gamma_{j}(u,u)}}$ and $K_{2}:=\sup_{\mathcal{C}}\sqrt{|\frac{1+\sqrt{1+\frac{|\rho(u,u)|^{2}}{|\tau_{j}|^{2}|k^{TT}_{j}(u,u)|^{2}}}}{2}|}\\\nonumber \sup_{\mathcal{C}}\sqrt{\frac{1+2\rho(u,u)}{2|\tau_{j}|\gamma_{j}(u,u)}\nonumber+\frac{\sqrt{1+2\mathcal{S}(\tau_{j})^{2}\tau^{2}_{j}}}{2|\tau_{j}|}}$. Under the assumption $|k^{TT}_{j}(u,u)|>\delta~\forall j$, we observe that $K_{1}\to \sqrt{\frac{a}{\sqrt{2}}}$ and $K_{2}\to \sqrt{\frac{a}{\sqrt{2}}}$ as $j\to\infty$.\\
\textbf{Remark:} Note that the limit of the ratio $\mu_{vert_{j}}(\mathcal{C})/l_{\gamma_{j}}(\mathcal{C})$ is $\sqrt{\frac{a}{\sqrt{2}}}$ which is a universal constant and therefore does not depend on the chosen sequence.\\

\section{Proof of theorem 2}
\noindent   {Recall the definition of the  measure of a closed curve $\mathcal{C}$ in a particular homotopy class with respect to the vertical foliation associated with the homolomorphic quadratic differential $\phi$ (or $k^{TT}$) 
\begin{eqnarray}
\mu_{vert}(\mathcal{C})&=&\oint_{\mathcal{C}}\sqrt{\Bigl|\frac{k^{TT}_{ij}u^{i}u^{j}+\sqrt{(k^{TT}_{ij}u^{i}u^{j})^{2}+(k^{TT}_{ij}u^{i}v^{j})^{2}}}{2}\Bigr|}d\lambda.
\end{eqnarray}
Let $\{\gamma_{j},k^{TT}_{j},\tau_{j}\}$ be any sequence such that $\tau_{j}\to-\infty$ as $j\to\infty$, that is, $\gamma_{j}$ leaves every compact set of $\mathcal{T}\Sigma_{p}$. Lemma \ref{theorem2proof3} yields 
\begin{eqnarray}
\lim_{j\to\infty}K_{1}\leq \lim_{j\to\infty}\frac{\mu_{vert_{j}}(\mathcal{C})}{l_{\gamma_{j}}(\mathcal{C})}\leq \lim_{j\to\infty}K_{2}    
\end{eqnarray}
or 
\begin{eqnarray}
\label{constancy}
\lim_{j\to\infty}\frac{\mu_{vert_{j}}(\mathcal{C})}{l_{\gamma_{j}}(\mathcal{C})}=\sqrt{\frac{a}{\sqrt{2}}}.    
\end{eqnarray}}

\noindent  {In addition to the transverse measure with respect to the vertical foliation, we also have the following transverse measure with respect to the horizontal foliation of the holomorphic quadratic differential $\phi$
\begin{eqnarray}
\mu_{hor}(\mathcal{C})&=&\oint_{\mathcal{C}}\sqrt{\Bigl|\frac{\sqrt{(k^{TT}_{ij}u^{i}u^{j})^{2}+(k^{TT}_{ij}u^{i}v^{j})^{2}}-k^{TT}_{ij}u^{i}u^{j}}{2}\Bigr|}d\lambda.
\end{eqnarray}
In the analysis of Wolf \cite{wolf1989teichmuller}, it is shown that this transverse measure associated with the horizontal foliation collapses asymptotically.
In Wolf's \cite{wolf1989teichmuller} construction, the domain is fixed while the target is varied, that is the dynamics occur in the target space. In my case, the dynamics take place in the domain. Therefore, we can not utilize the available machinery such as the Beltrami differential $\nu:=\frac{|W_{\bar{z}}|}{|W_{z}|}$ ($W: \Sigma_{p}(\gamma)\to \Sigma_{p}(\rho)$ and harmonic) or the associated Bochner equation controlling the behavior of $\nu$ to show that $\mu_{hor}$ vanishes and therefore, $k^{TT}_{ij}u^{i}v^{j}$ approaches zero asymptotically. Once again the Gauss map equation (\ref{eq:gauss}) together with the Lichnerowicz equation (relativistic version of the Bochner equation) come to the rescue and notably, they are of purely relativistic origin. Now consider a sequence $\{\gamma_{j},k^{TT}_{j},\tau_{j}\}$ such that $\tau_{j}\to-\infty$ as $j\to\infty$, that is, $\gamma_{j}$ leaves every compact set of $\mathcal{T}\Sigma_{p}$. Now upon contraction with $u$ and $v$, the Gauss-map equation yields
\begin{eqnarray}
|k^{TT}_{j}(u,v)|\leq \left(\frac{1}{2|\tau_{j}|}+\frac{\sqrt{1+2\mathcal{S}(\tau_{j})^{2}\tau^{2}_{j}}}{2|\tau_{j}|}\right)\gamma_{j}(u,v)+\frac{1}{|\tau_{j}|}\rho(u,v)\\\nonumber 
=\frac{1}{|\tau_{j}|}\rho(u,v)
\end{eqnarray}
since $\gamma_{j}(u,v)=0$ (\ref{eq:orth}). Therefore, we obtain 
\begin{eqnarray}
0\leq \mu_{hor_{j}}(\mathcal{C})\leq \sup_{\mathcal{C}}\sqrt{|\frac{\sqrt{1+\frac{|\rho(u,v)|^{2}}{|\tau_{j}|^{2}|k^{TT}_{j}(u,u)|^{2}}}-1}{2}|}\oint_{\mathcal{C}}\sqrt{|k^{TT}_{j}(u,u)|}   \\\nonumber
\leq \sup_{\mathcal{C}}\sqrt{|\frac{\sqrt{1+\frac{|\rho(u,v)|^{2}}{|\tau_{j}|^{2}|k^{TT}_{j}(u,u)|^{2}}}-1}{2}|}\sup_{\mathcal{C}}\sqrt{\frac{1+2\rho(u,u)}{2|\tau_{j}|\gamma_{j}(u,u)}+\frac{\sqrt{1+2\mathcal{S}(\tau_{j})^{2}\tau^{2}_{j}}}{2|\tau_{j}|}}\oint_{\mathcal{C}}\sqrt{\gamma_{j}(u,u)}
\end{eqnarray}
but with $|k^{TT}(u,u)|>\delta$ for any $\delta>0$,  $\sup_{\mathcal{C}}\sqrt{|\frac{\sqrt{1+\frac{|\rho(u,v)|^{2}}{|\tau_{j}|^{2}|k^{TT}_{j}(u,u)|^{2}}}-1}{2}|}\to 0$ and \\ $\sup_{\mathcal{C}}\sqrt{\frac{1+2\rho(u,u)}{2|\tau_{j}|\gamma_{j}(u,u)}+\frac{\sqrt{1+2\mathcal{S}(\tau_{j})^{2}\tau^{2}_{j}}}{2|\tau_{j}|}}\to \sqrt{\frac{a}{\sqrt{2}}}$ as $j\to\infty$. Therefore 
\begin{eqnarray}
\lim_{j\to\infty}\frac{\mu_{hor_{j}}(\mathcal{C})}{l_{\gamma_{j}}(\mathcal{C})}=0.    
\end{eqnarray}
Thus, the high Dirichlet energy limit (while viewed as a proper function on the Teichm\"uller space of the domain) precisely indicates that the transverse measure with respect to the horizontal foliation associated with the quadratic differential $\phi$ defined in terms of $k^{TT}$ (or equivalently $k^{TT}$) becomes negligible compared to the hyperbolic length (and hence the transverse measure to the vertical foliation by \ref{constancy}). This concludes the proof.} 

\section{ {Proof of theorem 3}}
\noindent Now we prove the final theorem that essentially indicates that the space of big-bang singularity is characterized by a subset of the Thurston boundary of the Teichm\"uller space. Let us consider the functional space $\Omega=\mathbb{R}^{G(\Sigma_{p})}_{>0}$, where the space of geodesics on $\Sigma_{p}$ is denoted as $G(\Sigma_{p})$, which may be obtained by the $\pi_{1}(\Sigma_{p})$ action on the space of geodesics on $\mathbb{H}^{2}$, that is $\mathbb{S}^{1}_{\infty}\times \mathbb{S}^{1}_{\infty}-\Delta$, $\Delta$ being the diagonal. Essentially $\Omega$ consists of functionals that take one element from each homotopy class of $G(\Sigma_{p})$ and associate a positive number to such element. It can essentially be viewed as the space of geodesic currents given that the association of a measure to $G(\Sigma_{p})$ is $\pi_{1}(\Sigma_{p})$ invariant (Radon measure to be precise). We may construct the following map 
\begin{eqnarray}
l: \mathcal{T}\Sigma_{p}\to \Omega\\
\mathcal{C}\mapsto l_{\mathcal{C}}: G(\Sigma_{p})\to R_{>0}.
\end{eqnarray}
We may projectivize the space $\Omega$ as follows
\begin{eqnarray}
\mathcal{P}\Omega&=&\Omega/(\beta\sim t\beta, t>0, \beta\in\Omega)
\end{eqnarray}
and subsequently, obtain the following  map
\begin{eqnarray}
\pi\circ l: \mathcal{T}\Sigma_{p}\to \mathcal{P}\Omega,
\end{eqnarray}
where $\pi$ is the canonical projection map $\pi:\Omega\to \mathcal{P}\Omega$.
Clearly, the map $l$ can be identified with the Liouville currents (see appendix A) (or the `$9g-9$' map). The injectivity of $\pi\circ l$ follows from the injectivity of the map $L$ of \ref{injectivity}.  
Similarly, we may construct the following map from the space of measured laminations to $\Omega$
\begin{eqnarray}
\nu: \mathcal{MF}\to \Omega\\\nonumber
\mathcal{F}\mapsto (i(\mathcal{F},\mathcal{C})=\oint_{[\mathcal{C}]\in G(\Sigma_{p})}\mu_{\mathcal{F}}).
\end{eqnarray}
Here, $\mu_{\mathcal{F}}$ corresponds to the transverse measure associated with $\mathcal{F}\in\mathcal{MF}$ and $[\mathcal{C}]$ is a homotopy class. Clearly, the space of measured geodesic foliations is a subset of the space of all geodesics and therefore, we have the following 
\begin{eqnarray}
\label{projectivedef}
\mathcal{PMF}:=(\mathcal{MF}-\{0\})/(\mathcal{F}\sim t\mathcal{F}, t>0, \mathcal{F}\in\mathcal{MF})\\\nonumber
=\pi\circ\nu(\mathcal{MF}) \subset \mathcal{P}\Omega.
\end{eqnarray}
Note that $\pi\circ\nu$ is injective. Let the space of holomorphic quadratic differentials with respect to the conformal structure $(M,\gamma)$ and $(M,\rho)$ be defined as $\mathcal{HQD}(\gamma)$ and $\mathcal{HQD}(\rho)$, respectively. Now we define another important map namely the Hubbard-Masur homeomorphism 
\begin{eqnarray}
\mathcal{F}: \mathcal{HQD}(\gamma)&\to& \mathcal{MF}\\
&\phi\mapsto&\mathcal{F}(\phi).
\end{eqnarray}
Recall that $\mathcal{PMF}$ and $\mathcal{T}\Sigma_{p}$ are disjoint in the space of projective currents i.e., $\mathcal{P}\Omega$ (see \ref{injectivity}). The Thurston compactification, essentially, is given as $\bar{\mathcal{T}\Sigma_{p}}^{Th}=\mathcal{T}\Sigma_{p}\cup \mathcal{PMF}$. $\mathcal{PMF}$ is the Thurston boundary of the Teichm\"uller space. We need the following final lemma to complete the proof of theorem 3.

\begin{lemma}
\label{lemmacrucial} \textit{Let the sequence $\{\gamma_{j}\}$ leave all the compact sets in $\mathcal{T}\Sigma_{p}$ at the limit of big-bang i.e., $j\to\infty$ ($\tau_{j}\to-\infty$). Then $\pi\circ l(\gamma_{j})$ converges if and only if $\pi\circ \nu (\mathcal{F}_{j})$ converges and subsequently both have the same limit in $\mathcal{P}\Omega$. Here $\mathcal{F}_{j}$ is the vertical measured foliation associated to the holomorphic quadratic differential $\phi_{j}=(k^{TT}_{11j}-i k^{TT}_{12j})dz^{2}$ through the Hubbard-Masur homeomorphism $\mathcal{F}$.}
\end{lemma}
\begin{proof} Note that the topology of $\mathcal{T}\Sigma_{p}$ is given by a finite number of curves $\mathcal{C}_{1},\mathcal{C}_{2},\mathcal{C}_{3},\cdot,\cdot,\cdot,\cdot,\mathcal{C}_{k}$ (e.g., $9g-9$ theorem). Suppose $\{\gamma_{j}\}$ diverges in $\mathcal{T}\Sigma_{p}$ and $\pi\circ l$ converges in $\mathcal{P}\Omega$. Then there exists a sequence of scalars $\lambda_{j}$ such that $\lambda_{j}\gamma_{j}$ converges. This means $\sum_{k}l_{\gamma_{j}}(\mathcal{C}_{k})\to\infty$ and $\sum_{k}\lambda_{j}l_{\gamma_{j}}(\mathcal{C}_{k})$ converges. Then lemma \ref{theorem2proof3} yields for $[\mathcal{C}]\in \pi_{1}(\Sigma_{p})$,
\begin{eqnarray}
|\lambda_{j}l_{\gamma_{j}}([\mathcal{C}])-\lambda_{j} \frac{2^{\frac{1}{2}}}{a^{\frac{1}{4}}}i(\mathcal{F}_{j},\mathcal{C})|\to 0
\end{eqnarray}
since $K_{1}\to \frac{a^{\frac{1}{2}}}{2^{\frac{1}{4}}}$ and $K_{2}\to \frac{a^{\frac{1}{2}}}{2^{\frac{1}{4}}}$ and $\gamma_{j}$ leaves every compact set of $\mathcal{T}\Sigma_{p}$. Here $a$ is a universal constant independent of the homotopy class of curves $[\mathcal{C}]$ and scaling measured foliation by a constant does not change the foliation only scales the measure. This makes no difference at the level of projective space. Thus we prove the one way. The other way may be obtained in a similar way. This concludes the proof of the lemma. 
\end{proof}

\noindent This lemma tells us that a sequence on solution curve (of Einstein's reduced equations) diverging (leaving every compact set) in the configuration space ($\mathcal{T}\Sigma_{p}$), converges in the space of projective measured foliations as $\tau_{j}\to-\infty$ (big-bang limit) since by definition \ref{projectivedef}, the convergent limit of $\pi\circ \nu(\mathcal{F}_{j})$ lies in $\mathcal{PMF}$. The space of projective measured foliations $\mathcal{PMF}$ is well known to be the Thurston boundary of the Teichm\"uller space. Therefore the limit of $\pi\circ l(\gamma_{j})$ lies on the Thurston boundary by lemma \ref{lemmacrucial}. This establishes theorem 3. 

\begin{remark}
Note that this \textit{does not} prove the reverse i.e., and every boundary point of the Thurston compactified Teichm\"uller space is the limit of a unique
ray. In other words, we can only infer at present that the space of big-bang singularity is characterized by a subset of the Thurston boundary of the Teichm\"uller space. However, we conjecture the following
\noindent \textbf{Conjecture:}
\textit{Thurston boundary of the Teichm\"uller space is exactly the space of big-bang singularities}.   
    
\end{remark}

\section{Concluding Remarks}
\noindent Despite the fact that `2+1' gravity is devoid of a straightforward physical significance due to the lack of gravitational wave degrees of freedom, it is of extreme importance while studying `3+1' gravity on spacetimes of certain topological type ($\mathbb{S}^{2}\times \mathbb{S}^{1}\times \mathbb{R},  \mathbb{T}^{2}\times \mathbb{S}^{1}\times \mathbb{R}$, and $\Sigma_{p}\times \mathbb{S}^{1}\times \mathbb{R}$, non-trivial $\mathbb{S}^{1}$ bundles over $\Sigma_{p}\times \mathbb{R}$). As mentioned in the introduction, several studies have been done on this topic where the `$3+1$' gravity has been realized as the $`2+1$' gravity coupled to a wave map, and where the Teichm\"uller space of $\Sigma_{p}$ plays a crucial role. In the $2+1$ case, the configuration space is the Teichm\"uller space and we have shown here that the space of big-bang singularities is realized as a subset of the Thurston boundary of the Teichm\"uller space. At the big bang, the conformal geometry degenerates via pinching and wringing of $(\Sigma_{p},\gamma)$. This result essentially characterizes the complete solution space as well as indicates that the reduced Einstein flow can potentially be used to compactify Teichm\"uller space if one is able to prove the second part of the question \textit{every boundary point of the Thurston compactified Teichm\"uller space is the limit of a unique
ray}. While my result is obtained by studying purely vacuum gravity, a natural question arises whether the inclusion of a positive cosmological constant might yield the same result. \cite{anderson1997global, moncrief2019could, fajman2020future} studied vacuum GR with a positive cosmological constant in $2+1$ case, where the future in time behavior seems to persist. Therefore, it would be interesting to include a positive cosmological constant and check whether the Thurston boundary is approached in the big-bang limit. In addition, if one includes matter source and focuses on the evolution of the gravitational degrees of freedom (due to the presence of matter sources, the configuration space is now infinite dimensional), can the big-bang limit be realized as a subset of the Thurston boundary? Could the Teichm\"uller degrees of freedom of `3+1' gravity on $U(1)$ symmetric $\mathbb{S}^{1}$ bundles over $\Sigma_{p}\times \mathbb{R}$ realize a subset of the Thurston boundary in the same limit? What is the implication of such limiting behavior at the classical level in quantizing `2+1' gravity or `3+1' gravity on these special topologies? Can this characterization of the space of singularities be extended to higher dimensional gravity? Can the Einstein flow be used further to study classical Teichm\"uller theory?            
\begin{center}
\begin{figure}
\begin{center}
\includegraphics[scale=0.4]{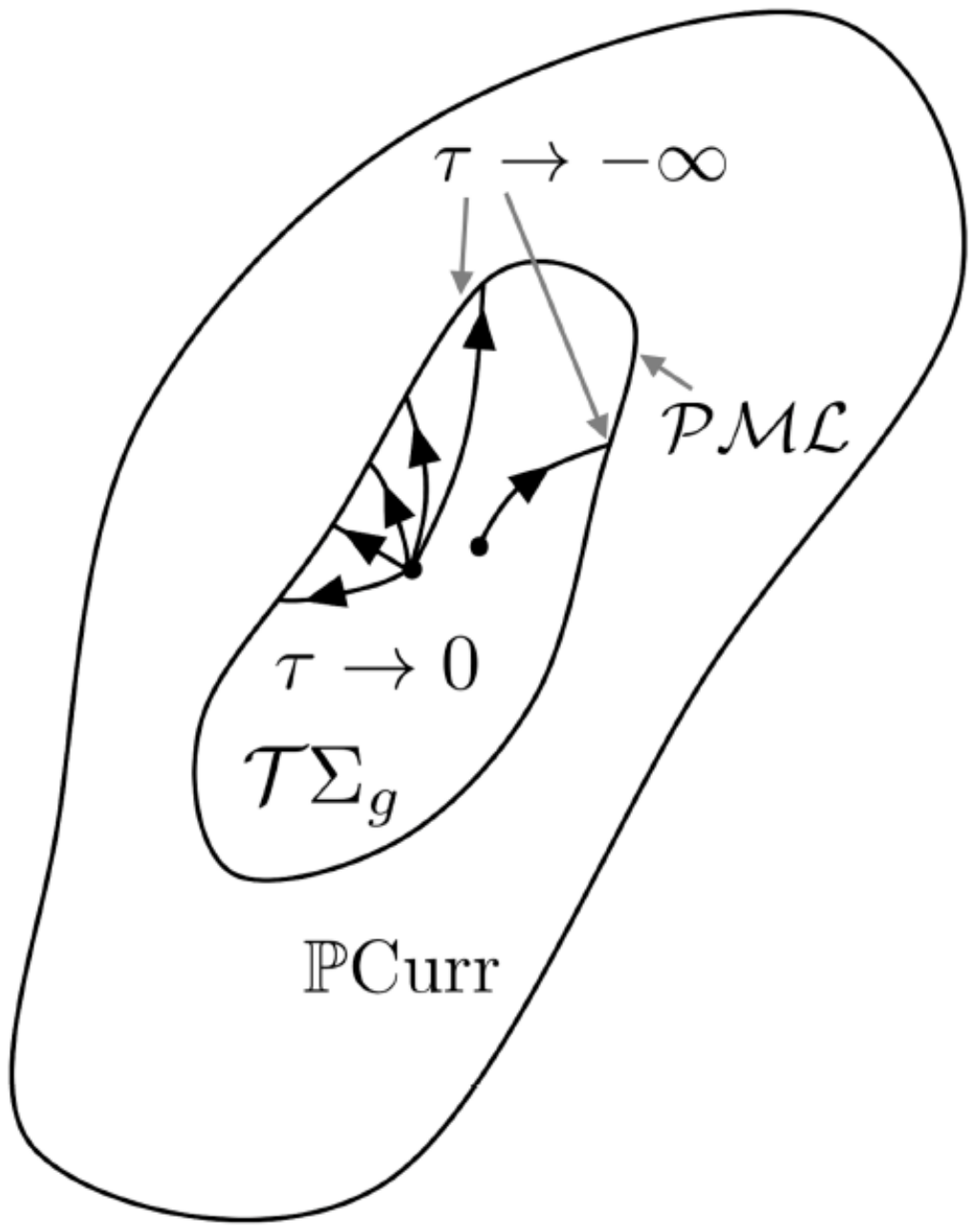}
\end{center}
\begin{center}
\caption{The schematics of the reduced dynamics on the configuration space $\mathcal{T}\Sigma_{p}$ ($\approx \mathbb{R}^{6p-6}$). Each solution curve starts at $\tau\to0$ and approaches the Thurston boundary of $\mathcal{T}\Sigma_{p}$ in $\mathbb{P}$Curr (or equivalently $\mathbb
{P}\Omega$) in the limit of big-bang.}
\label{fig:pdf}
\end{center}
\end{figure}
\end{center}

\section{Acknowledgement}
\noindent I would like to thank Prof. Vincent Moncrief and Prof. Shing-Tung Yau for numerous useful discussions related to this project and their help in improving the manuscript. I would also like to thank the referee for a very thorough review. This work was supported by Yale University and CMSA at Harvard University.

\begin{appendix}
\label{AA}
\section{Space of projective laminations as the Thurston boundary of the Teichm\"uller space}
\noindent In this section, we provide a rough sketch of the proof of Thurston compactification of the Teichm\"uller space by the space of projective measured laminations ($\mathcal{PML}$). The details may be found in \cite{fathi2012thurston, mcmullen2011riemann}. Here we show that a sequence diverging in Teichm\"uller space converges in the space of projective measured laminations ($\mathcal{PML}$), which is a compact subset of the space of geodesic currents. A  $\pi_{1}(\Sigma_{p})-$invariant measure on $G(\tilde{\Sigma}_{g})$ may be defined as
\begin{eqnarray}
\hat{L}=\frac{d\alpha d\beta}{|e^{i\alpha}-e^{i\beta}|^{2}},
\end{eqnarray}
where $(e^{i\alpha},e^{i\beta})\in(\mathbb{S}^{1}\times \mathbb{S}^{1})\setminus \Delta$ and $\Delta$ represents diagonal. This measure is called the Liouville measure.The Liouville measure corresponding to $X$ is denoted by $\hat{L}_{X}$, which satisfies the following for any $\gamma\in G(\Sigma_{p})$
\begin{eqnarray}
i(\gamma,\hat{L}_{X})=l_{X}(\gamma),
\end{eqnarray}
where $i$ denotes the bilinear function 'intersection number' and $l_{X}(\gamma)$ denotes the length of $\gamma$ with respect to the hyperbolic metric on $X$. 
The intersection property may be interpreted as follows. Let us consider a closed non-trivial geodesic $\gamma\in G(\Sigma_{p})$. Lift $\gamma$ to the universal cover and consider its intersection with the set of geodesics transverse to its lift $\tilde{\gamma}$ that is $i(\gamma,\hat{L})$ is defined as $\int_{E}\hat{L}(E\cap \tilde{\gamma})$, where $E\subset G(\tilde{\Sigma}_{g})$ is the set of geodesics transverse to $\tilde{\gamma}$. A few lines of calculations show that this integral is indeed the length of $\gamma$ with respect to the hyperbolic metric (scalar curvature =-1). Note that the Liouville measure may be used to define geodesic currents on $G(\Sigma_{p})$ due to its $\pi_{1}(\Sigma_{p})-$invariance property. Now, let $(X,f)$ be a hyperbolic surface (and thus $\in \mathcal{T}\Sigma$) such that $f: \Sigma_{p}\to X=\mathbb{H}^{2}/\pi_{1}(\Sigma_{p})$ is a homeomorphism. Liouville measure provides a well-defined map from the Teichm\"uller space $\mathcal{T}\Sigma$ to the space of currents. Here we just provide a brief description of the Thurston compactification of the Teichm\"uller space, necessary for the current purpose. For details, the readers are referred to the excellent book \cite{fathi2012thurston}, where the proof of the stated theorems may be found.\\
\begin{lemma}
\cite{fathi2012thurston, mcmullen2011riemann} \textit{The map $(X,f)\to \hat{L}_{X}$ is a proper embedding of $\mathcal{T}\Sigma_{p}$ into the space of currents Curr$(\Sigma_{p})$ given by the intersection number $i$ that is, for all closed curves $\alpha$ in $\Sigma_{p}$,
\begin{eqnarray}
\label{eq:image}
i(\alpha,\hat{L}_{X})=l_{X}(\alpha)
\end{eqnarray}}
defines a proper embedding of $\mathcal{T}\Sigma_{p}$ into Curr$(\Sigma_{p})$.
\end{lemma} 
\begin{proof} See \cite{fathi2012thurston, mcmullen2011riemann}.
\end{proof}
We are now ready to establish the Thurston compactification. Let us first state a lemma.\\
\begin{lemma}  \cite{fathi2012thurston, mcmullen2011riemann} For any hyperbolic surface $\Sigma_{p}$ with the marking $(X,f)$, we have the following result
\begin{eqnarray}
i(\hat{L}_{X},\hat{L}_{X})=\pi^{2}|\chi(\Sigma_{p})|,
\end{eqnarray} 
where $\chi(\Sigma_{p})=2(1-g))$ is the Euler characteristics of $\Sigma_{p}$.
\end{lemma}
Remarkably, this is a topological invariant. Let's denote the map $(X,f)\to \hat{L}_{X}$ by $\hat{L}$. We have the following lemma
\begin{lemma}
\label{injectivity}
\begin{eqnarray}
\hat{L}: \mathcal{T}\Sigma_{p}\to \rm I\! PCurr(\Sigma_{p})=(Curr(\Sigma_{p})-0)/(\mu\sim t\mu,\mu\in Curr (\Sigma_{p}), t\in \rm I\! R_{>0})\nonumber
\end{eqnarray} 
is injective.
\end{lemma}
\begin{proof}
Let $[f: \Sigma_{p}\to X]$ and $[h:\Sigma_{p}\to Y]$ be two elements of $\mathcal{T}\Sigma_{p}$. Then 
\begin{eqnarray}
[\hat{L}_{X}]=[\hat{L}_{Y}]=>\hat{L}_{X}=t\hat{L}_{Y}.
\end{eqnarray}
Now we use the previous lemma and obtain
\begin{eqnarray}
\pi^{2}|\chi(\Sigma_{p})|=i(\hat{L}_{X},\hat{L}_{X})=i(t\hat{L}_{Y},t\hat{L}_{Y})\\\nonumber
=t^{2}i(\hat{L}_{Y},\hat{L}_{Y})=t^{2}\pi^{2}|\chi(\Sigma_{p})|,
\end{eqnarray}
i.e., 
\begin{eqnarray}
t=1,
\end{eqnarray}
as $t\in \rm I\! R_{>0}$ and therefore $L_{X}=L_{Y}$. 
\end{proof}
As we have defined earlier, a lamination $\mathcal{L}$ on $\Sigma_{p}$ is a closed subset which is the union of disjoint simple geodesics, and the geodesics in $\mathcal{L}$ are called the leaves of the lamination. An important property of these leaves is that they do not intersect each other that is if $\lambda, \alpha\in \mathcal{L}$, then the following is satisfied 
\begin{eqnarray}
i(\lambda,\alpha)=0.
\end{eqnarray}
If we associate a transverse measure to the leaves of $\mathcal{L}$, then we obtain a measured lamination denoted by $\mathcal{ML}$. we may of course construct the projective measured laminations $\mathcal{PML}$ through the following identification 
\begin{eqnarray}
\mathcal{PML}=(\mathcal{ML}-\{0\})/(\lambda\sim t\lambda, \lambda\in \mathcal{ML}, t>0).
\end{eqnarray}
Clearly, the leaves of a measured lamination define a subset in the space of all geodesics, and therefore, the projective measured lamination $\mathcal{PML}$ may be identified as a subset of the space of geodesic currents. It is in fact a compact subset, which may be proven utilizing an elementary result from topology namely Tychonof's theorem \cite{bmybaki1995elements}. Another important observation is to note that the image of the Teichm\"uller space under the map $\hat{L}$ i.e., $\hat{L}(\mathcal{T}\Sigma_{p})$ and $\mathcal{PML}$ are disjoint. This follows from the definition of the geodesic lamination that is $i(\lambda,\lambda)=0~\forall\lambda\in \mathcal{PML},$ while $i(\hat{L}_{X},\hat{L}_{X})=\pi^{2}|\chi(\Sigma_{p})|\neq 0,~\forall X\in\mathcal{T}\Sigma_{p}.$
Now we finish the Thurston compactification \\
\begin{lemma} The closure of $\mathcal{T}\Sigma_{p}\subset \rm I\! PCurr(\Sigma)$ is precisely $\mathcal{T}\Sigma_{p}\cup\rm I\! P\mathcal{ML}$.
\end{lemma}
\begin{proof} Let say $[f_{n}: \Sigma_{p}\to X_{n}]$ is a sequence that diverges in $\mathcal{T}\Sigma_{p}$. Then obviously, $\{[\hat{L}_{X_{n}}]\}\subset \mathcal{P}$Curr$(\Sigma_{p})$ converges to some element of $\mathcal{P}$Curr$(\Sigma_{p})$ due to the fact that $\mathcal{P}$Curr$(\Sigma_{p})$ is a compact subset of Curr$(\Sigma_{p})$ (passing to a subsequence). Then $\exists~t_{n}$ such that 
$Lim_{n\to\infty}t_{n}\hat{L}_{X_{n}}=\mu\in $PCurr$(\Sigma_{p}).
$
Now from the divergence criteria, there exists a simple closed curve $\alpha\in \Sigma_{p}$, such that 
\begin{eqnarray}
Lim_{n\to\infty}l_{X_{n}}(\alpha)=\infty.
\end{eqnarray} 
But, $\infty>i(\alpha,\mu)=i(\alpha,t_{n}\hat{L}_{X_{n}})=t_{n}l_{X_{n}}(\alpha)$ and thus we must have 
\begin{eqnarray}
Lim_{n\to\infty}t_{n}=0.
\end{eqnarray}
Now we see the following 
\begin{eqnarray}
i(\mu,\mu)&=&i(lim_{n\to\infty}t_{n}\hat{L}_{X_{n}},lim_{n\to\infty}t_{n}\hat{L}_{X_{n}}),\\
&=&lim_{n\to\infty}t_{n}^{2}i(\hat{L}_{X_{n}},\hat{L}_{X_{n}}),\\
&=&lim_{n\to\infty}t_{n}^{2}\pi^{2}|\chi(\Sigma)|,\\
&=&0,
\end{eqnarray}
and therefore, $\mu\in\mathcal{PML}$.
\end{proof}

\section{Approaching $\partial \mathcal{T}\Sigma_{p}$}
\noindent Let us consider the Fenchel Neilsen coordinates of the Teichm\"uller space. Figure (\ref{fig:fenchel}) shows the pants decomposition of the Teichm\"uller space and the associated Fenchel-Neilen co-ordinates (see \cite{FM11} for the details of the Fenchel-Neilsen parametrization and pants decomposition).  Such parametrization is given by the lengths of $3g-3$ nontrivial (nontrivial in $\pi_{1}(\Sigma_{p})$) geodesics $\{l_{i}\}_{i=1}^{3g-3}$ along with $3g-3$ associated twist parameters $\{\theta_{i=1}^{3g-3}\}$ (twist is performed about the same geodesic). The two possible mechanisms of attaining the boundary of the Teichm\"uller space are described below.

\subsection{Pinching of $\Sigma$}    
\noindent Let $\gamma(l_{i}^{n})$ denotes a sequence of hyperbolic metrics and let $\theta_{i}=0~\forall i=1,2,3,....,3g-3$. Letting any one of the $l_{i}$ tend to infinity i.e., $\lim_{n\to\infty}l_{i}^{n}=\infty$ implies approaching the boundary $\partial\mathcal{T}\Sigma$. Using the collar lemma (see \cite{tromba2012teichmuller} for the detailed proof of the collar lemma), we immediately obtain that there is a non-trivial geodesic transverse to $l^{n}_{i}$ with length $\approx \lim_{n\to\infty}e^{-l_{i}^{n}}$. Note that the nontrivial (in $\pi_{1}(\Sigma_{p})$) geodesic $\mathcal{\gamma}_{2}$ collapses while the hyperbolic length $l_{1}$ of $\mathcal{\gamma}_{1}$ approaches infinity. Now, the Dirichlet energy of the harmonic map $id: (\Sigma_{p},\gamma(l^{n}_{i})) \to (\Sigma_{p},\rho)$ i.e., between fixed domain (with metric $\rho$) and the varying target (with metric $\gamma(l_{i})$) defined  is a continuous proper function on the Teichm\"uller space. Therefore, the sequence of Dirichlet energies associated with the diverging sequence of metrics (or degenerating to be precise) $\gamma(l^{n}_{i})$ can not stay in a compact set; that is the sequence blows up. Therefore we have the following correspondence 
\begin{eqnarray}
\lim_{n\to\infty}l^{n}_{i}\to\infty=> \lim_{n\to\infty}E_{\gamma(l^{n}_{i})}\to\infty.
\end{eqnarray}
 Notice that multiple non-trivial geodesics $\mathcal{\gamma}_{i}$ (and the corresponding transverse ones) may show the pinching behavior at once and each such limit corresponds to distinct points on $\partial\mathcal{T}\Sigma_{p}$. 
 
\subsection{Wringing of $\Sigma_{p}$ by its neck} 
\noindent In order to explain the approach to $\partial \mathcal{T}\Sigma_{p}$ through wringing of $\Sigma_{p}$, we need to introduce the symplectic geometry of the Teichmuller space \cite{wolpert2010families, wolpert2009weil}. Using the parametrization $(l_{i},\theta_{i})_{i=1}^{3g-3}$ of Teichmuller space, define the symplectic form
\begin{eqnarray}
\omega&=&\sum_{i=1}^{3g-3}dl_{i}\wedge d\theta_{i},
\end{eqnarray}    
which is preserved under the flow of the vector field $v=-\frac{\partial}{\partial\theta_{i}}$ and satisfies 
\begin{eqnarray}
\omega(-\frac{\partial}{\partial\theta_{i}},\cdot)&=&dl_{i}.
\end{eqnarray}
The conserved Hamiltonian is nothing but the length $l_{i}$. Here, $\theta_{i}$ is the twist parameter about the $i$th geodesic.
Therefore, flow of the vector field $-\frac{\partial}{\partial\theta_{i}}$ preserves the length $l_{i}$ of the geodesic about which $\Sigma_{p}$ is twisted. After $n$ such twists for large $n$, the length of the geodesic transverse to the $i$th geodesic increases by $\sim nl_{i}$. The wringing of $\Sigma_{p}$ about the $i$th geodesic corresponds to the limit $n\to\infty$. Let the length of the transverse geodesic before the twist be $L^{T}$. After performing $n$ twists, the length becomes $\sim L^{T}+nl_{i}$ and therefore, the wringing corresponds to the fact that $\lim_{n\to\infty}\frac{l_{i}}{L_{T}+nl_{i}}=0$. This is the other mechanism to approach the boundary of the Teichm\"uller space. Note that every point on the boundary $\partial\mathcal{T}\Sigma_{p}$ can be obtained through a combination of these two basic operations and in every situation, the Dirichlet energy approaches infinity.

\end{appendix}

\end{document}